\documentclass[aps,pre,groupedaddress,showkeys]{revtex4-2}

\usepackage[utf8]{inputenc}
\usepackage{amsmath}
\usepackage{mathtools}
\usepackage{amsfonts}
\usepackage{amssymb}
\usepackage{graphicx}
\usepackage{color}
\usepackage{multirow}
\usepackage{float}
\usepackage{array}
\usepackage{svg}
\usepackage{lineno}
\usepackage{hyperref}
\usepackage{natbib}
\hypersetup{
    colorlinks=true,
    linkcolor=blue,
    filecolor=blue,      
    urlcolor=blue,
    citecolor=blue,
}
\usepackage{array}
\newcolumntype{P}[1]{>{\centering\arraybackslash}p{#1}}
\usepackage[T1]{fontenc}
\newcommand*{\textcal}[1]{\textit{\fontfamily{qzc}\selectfont#1}}

\begin{document}


\title{Thixotropy in Viscoplastic Drop Impact on Thin Films}


\author{Samya Sen}
\author{Anthony G. Morales}
\author{Randy H. Ewoldt}\email{ewoldt@illinois.edu}
\affiliation{Department of Mechanical Science and Engineering\\University of Illinois at Urbana-Champaign\\Urbana, IL 61801, USA}


\date{\today}

\begin{abstract}

We use high-speed imaging to study the effect of thixotropic aging in drop impact of yield-stress fluids on pre-coated substrates. 
Our results reveal that 
drop splashing is suppressed for ``aged'' compared to ``unaged'' samples, indicating that thixotropic breakdown timescales during impact are long enough to affect the dynamics. 
We propose and test several hypotheses for modifying the dimensionless group $\text{IF}(D/t)$~\cite{BCB_PhysFluids2015,SSRHE_JFM2020} to account for thixotropic aging. 
The main challenge is that the steady flow properties (Herschel-Bulkley model parameters) used in the current dimensionless group cannot be defined or measured for thixotropically aged samples, because any deformation inherently rejuvenates and breaks down the microstructure. 
We find the most suitable hypothesis is to only increase the yield stress ($\sigma_{\rm y}$, plastic component) based on the storage modulus of aged samples, while keeping the viscous parameters ($K$ or $\eta_{\infty}$) constant. 
The work reveals fundamental insight into rarely studied short-timescale flow conditions with thixotropic effects. 
These results are important for applications such as fire suppression or spray coating that involve complex fluids of varying degrees of thixotropic aging.

\end{abstract}


\maketitle

\section{Introduction\label{sec:introduction}}
Yield-stress fluids flow only for a sufficiently large applied stress, and are otherwise predominantly solid \cite{dpl,NguyenBogerReview1992,BalmforthReview2014, CoussotReview2014, BonnManneville2017}. This non-Newtonian behavior enables diverse applications, including fire suppression, jet and spray printing, painting, coatings,  and direct-write 3D printing \cite{LuuForterre2009,Gunasekaran2009,BCBPhDThesis,AZN_SM2017,AZN_CurrOpin2019}. Such applications have motivated research in areas of drop, jet, and spray impacts of these complex fluids. Extensive literature exists for such tests with Newtonian fluids \cite{ReinJFM1996,CossaliMarengo1997,Wang_PhysFluids2000,SivakumarTropea2002,JosserandZaleski2003,VanderWalMozes2005,VanderWalMozes2005a}, but studies for non-Newtonian fluids are more limited, especially yield-stress fluids (also known as viscoplastic fluids) which are different both qualitatively and quantitatively \cite{Nigen2005,LuuForterre2009,Lohse_SM2012,BCB_PhysFluids2015,BCB_JNNFM2016,BCB_AtomSpray2017}. No studies have ever considered thixotropic aging effects, as we do here.

Researchers have previously studied the impact of various shear-thinning and yield-stress fluid drops on dry surfaces \citep{LuuForterre2009,Bertola2009,OishiJFM2019} and proposed scaling laws predicting different outcomes of the impacts (stick, spread and retraction, post-impact drop shape). The yield-stress fluids tested were mainly aqueous suspensions of clays such as Kaolin and Bentonite and soft glasses such as Carbopol. More weakly shear-thinning fluids (those without an apparent yield stress) tested were formulations of xanthan gum in water \citep{Bertola2009}. Some of these scaling laws have also been verified experimentally for Carbopol \citep{Lohse2012SM}, including water-entry problems \citep{JalaalLohse_JFM2019}.

Experiments studying yield-stress fluid drop impacts on films, relevant to spray coating and sequential drop impact, have only recently been reported \cite{LuuForterre2009, Bertola2009,BCB_PhysFluids2015, BCB_JNNFM2016, ChenBertola2017, BCB_AtomSpray2017, SSRHE_JFM2020}. Blackwell \textit{et al.}\ \cite{BCB_PhysFluids2015} tested drops of aqueous suspensions of polymeric microgel particles (Carbopol 940) impacting horizontal substrates coated with the same material and proposed an empirical non-dimensional scaling law for predicting stick-splash regimes taking into account the most dominant forces (inertial and dissipative flow effects). Yield-stress fluids can stick and stay localized at the point of impact, but the high momentum of the impacting drop can result in the formation of an ejection sheet, which can break and spatter away from the surface. Some, none, or all of these phenomena may be observed depending on the fluid properties and dynamic conditions. One thus needs an understanding of what influences the stick, splash, or spatter behavior of yield-stress fluid drops impacting coatings, focusing mainly on the fluid dynamic conditions and fluid flow and rheological properties. These results are crucial for applications such as spray coating and fire suppression where fluids are deposited in layers, and droplets interact with layers of films that have been sitting quiescently from an earlier deposition.

 Although short ($<1$~s) thixotropic timescales are often neglected because they are difficult to experimentally measure, such short timescales are relevant for droplet impact events, especially for impacts with very small droplets and at high velocities. Drop impacts under such conditions can have characteristic strain rates of $\mathcal{O}(1000)~\rm{s}^{-1}$, or corresponding timescales of $\mathcal{O}(1)$~ms. As a result, even very short thixotropic timescales become relevant and can influence drop impact behavior. Prior ideas and experiments with Carbopol were thus extended to drops and films of a yield-stress fluid with a completely different microstructure: Laponite RD, a clay suspension in water, by the authors \cite{SSRHE_JFM2020}. These two fluids, although possessing a yield stress, are vastly different in microstructure: Carbopol is a soft, jammed glassy system, and Laponite is a clay made of an attractive particulate network. The biggest difference is that Carbopol is marginally thixotropic, while Laponite is an appreciably thixotropic material, and drop impact scenarios with thixotropic effects have never been studied before. Earlier, studies on Laponite were limited to unaged or fresh samples to avoid the complicating effects of thixotropic aging, and it was shown that the dimensionless scaling worked well for these two distinct yield-stress systems \cite{SSRHE_JFM2020}. We address the thixotropic aspect of these fluids in the current work.

Thixotropy refers to time-dependent structure breakdown when subjected to a fixed shear rate, and subsequent time-dependent buildup at rest \cite{MewisReview1979, MewisWagnerReview2009}. Laponite is thixotropic, because the existence of a particulate network inside the solvent renders a structure to its suspension (schematically shown in Fig.~\ref{fig:setup} (A)). Increasing shear rate can breakdown this structure over time, and consequently the viscosity of the material decreases. The structure can buildup again upon cessation of shear (or stepping down to a lower shear rate). The buildup of structure with time is often referred to as aging, while the breakdown of structure is called rejuvenation. Many viscoplastic systems  exhibit thixotropy, especially those of a colloidal nature (like most clays, including Laponite) that have a structure to the fluid. Dimensionless groups used in drop impact studies, especially considering colloidal systems, must be modified to accommodate for this phenomenon, especially since such thixotropic aging can render established scaling relations invalid.

Hence, in this work, we aim to experimentally quantify the drop impact behavior of Laponite suspensions as a function of aging time, using the setup shown in Fig.~\ref{fig:setup}. We expect that aging time is important, but also  thixotropic \emph{breakdown time}, because the timescales associated with the drop impact events ($\mathcal{O}(10)$~ms) are either comparable to or shorter than the breakdown timescales of aged Laponite ($\mathcal{O}(100)$~ms, see Sec.~3, supplementary information). The influence of breakdown kinetics is discussed in more detail in Sec.~\ref{sec:results}.

In addition to reporting a thorough experimental data set, we study and extend a proposed dimensionless group \cite{BCB_PhysFluids2015,SSRHE_JFM2020} for predicting stick-or-splash regimes of viscoplastic fluids with appreciable thixotropy. This dimensionless group is a comparison between inertial forces and total dissipative flow forces, which includes both yield stress and viscous rate-dependent effects, expressed as \cite{SouzaMendes2004}

\begin{align}\label{eq:IFintro}
\text{IF} \left( \frac{D}{t} \right) \equiv \frac{\rho V^2}{\sigma_{\rm y} + K(V/t)^{0.5} + \eta_{\infty} V/t} \left( \frac{D}{t} \right),
\end{align}
where the shear yield stress $\sigma_{\rm y}$, consistency index $K$, and the infinite-shear viscosity $\eta_{\infty}$, are model parameters from the generalized Herschel-Bulkley viscoplastic model for steady flow. $D$ is the drop diameter prior to impact, $t$ is the film thickness, $\rho$ is the fluid density, and $V$ is the velocity of impact. The group IF is a ratio of stresses, while $\text{IF}(D/t)$ is a ratio of forces, based on the hypotheses of the different characteristic length scales involved \cite{BCB_PhysFluids2015}. Detailed explanation of this scaling is provided in our earlier works \cite{BCB_PhysFluids2015,SSRHE_JFM2020}. We test the efficacy of this group in multiple ways, namely: (i) if it is capable of separating impact behavior into distinct regimes, (ii) if this gives a constant critical value for a splash transition for Laponite, (iii) if this critical value is the same for all concentrations of the fluid, and, most importantly, (iv) whether thixotropic aging influences this critical value. We find that these criterion fail, unless Eq.~\ref{eq:IFintro} is modified to account for thixotropic aging. We propose and test multiple hypotheses for modifying Eq.~\ref{eq:IFintro}. We find that the most credible is to increase the effective value of $\sigma_{\rm y}$ based on the aged linear elastic modulus $G^{\prime}$, but keeping viscous parameters $K$ and $\eta_{\infty}$ fixed at ``rejuvenated'' values. With this, the modified Eq.~\ref{eq:IFintro} gives very similar critical values for both aged and unaged samples.

\section{Materials and methods\label{sec:matmeth}}

\subsection{Drop impact setup\label{subsec:methods}}

\begin{figure}
\centering
\includegraphics[scale=0.13]{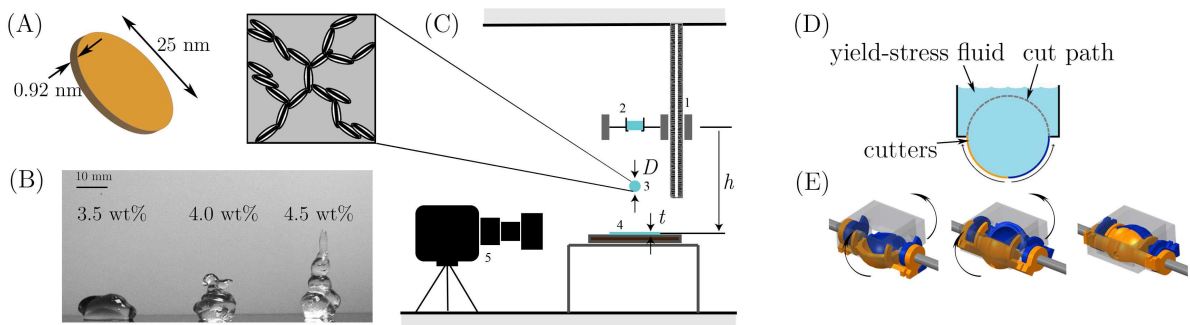}
\caption{\label{fig:setup}Experimental setup (figure reused with permission from \cite{SSRHE_JFM2020}). (A) Structure of Laponite clay particles, with disc-like particles stacked in suspension (image adapted from \cite{TanakaBonn2004}, reused with permission). (B) The yield stress of Laponite helps it retain its shape against gravity and this becomes more prominent as the concentration increases from 3.5 to 4.5 wt\%. (C) Schematic of drop impact setup. The geometric parameters: drop diameter, $D$, coating thickness, $t$, dropping height, $h$, are labeled. The components are: 1 - variable height trolley, 2 - drop maker with fluid, 3 - discharged drop, 4 - fluid coating on substrate, 5 - high-speed camera. (D) The schematic of the drop maker and the side view of how the spherical shape is formed. (E) 3-D model of the device in starting, intermediate, and final positions (left-to-right).}
\end{figure}

Fig.~\ref{fig:setup} outlines the experimental setup. It is the same arrangement used in \cite{SSRHE_JFM2020}, which can be referred to for exact details. Briefly, spherical drops of different diameters formed with the help of a special ``drop-maker'' device were impacted onto films of various thicknesses at varying velocities. The tests were repeated with different Laponite concentrations and both states of thixotropic aging (unaged and aged). The experimental conditions and steady-shear rheological fit parameters (for use in Eq.~\ref{eq:IFintro}) are listed in Table \ref{table:params_unaged}. A pair of samples of each concentration, one aged ($\tau_{\text{age}} = 600$~s) and the other unaged, were tested for each experimental condition. For each combination of parameters and sample aging state, duplicate tests were done to assess repeatability, which amounted to 2,016 impact tests. After each test, for the unaged samples, the coating was removed and replaced with a freshly stirred sample, identical to the one loaded on the drop maker, and the tests conducted within a 10~s window. To test the aged counterpart, tests were conducted after freshly preparing the film and aging both the drop and film for $\tau_{\text{age}} = 10$~min.

\begin{table}[tb]
\caption{\label{table:params_unaged}Range of experimental (geometric and material) parameters explored. The ranges of the quantities $\sigma_{\rm y}$, $K$, and $\eta_{\infty}$ (steady-state flow) are determined by the range of weight fractions of Laponite tested.}
  \begin{center}
  \begin{tabular}{p{3.5cm}p{1.2cm}p{1.2cm}|p{3.5cm}p{1.2cm}p{1.2cm}}
      fluid parameter   &   min   &   max  \;\;  & \;\; test parameter   &   min   &   max  \\[3pt]
      \hline
	$\sigma_{\rm y}$ [Pa]  & 39 & 67 \;\;&\;\;\; $V$ [m s$^{-1}$] & 2 & 6\\
	$K$ [Pa$\cdot$s$^{0.5}$] & 0.04 & 0.95 \;\;& \;\; $D$ [mm] & 10 & 20\\
	$\eta_{\infty}$ [Pa$\cdot$s] & 0.02 & 0.03 \;\;&\;\;\, $t$ [mm] & 0.25 & 3.18\\
  \end{tabular}
  \end{center}
\end{table}

\subsection{Materials and rheological characterization\label{subsec:materials}}

Laponite (Fig.~\ref{fig:setup} (A), (B)) is a crystalline powder of colloidal discs composed of lithium sodium magnesium silicate ($\text{Na}^+_{\text{0.7}}\left[\text{Si}_{\text{8}}\text{Mg}_{\text{5.5}}\text{Li}_{\text{0.3}}\text{O}_{\text{20}}(\text{OH})_{\text{4}}\right]^-_{\text{0.7}}$), and forms a soft gel when dispersed in aqueous solution. The platelet-shaped particles are approximately 25~nm in diameter and 0.92~nm thick \cite{Cummins2007}. The faces of the discs are negatively charged, and depending on the pH of the solution, the sides can hold partial positive charge \cite{Bonn1999}. Due to this, attractive forces exist between the discs and they come together to form a stacked house of cards structure \cite{Bonn1999,TanakaBonn2004,Cummins2007} (Fig.~\ref{fig:setup} (A)). The suspensions used in the experiments were prepared following the protocol outlined in \cite{SSRHE_JFM2020}. Samples tested within 10~s of stirring were deemed as ``unaged'' (minimal thixotropic restructuring), while those tested after 10~min were called ``aged'' \cite{MewisReview1979,MewisWagnerReview2009}.

\begin{figure}[!ht]
\begin{minipage}{0.49\textwidth}
\centering
    \includegraphics[scale=0.4,trim={0 0 0 0.5cm},clip]{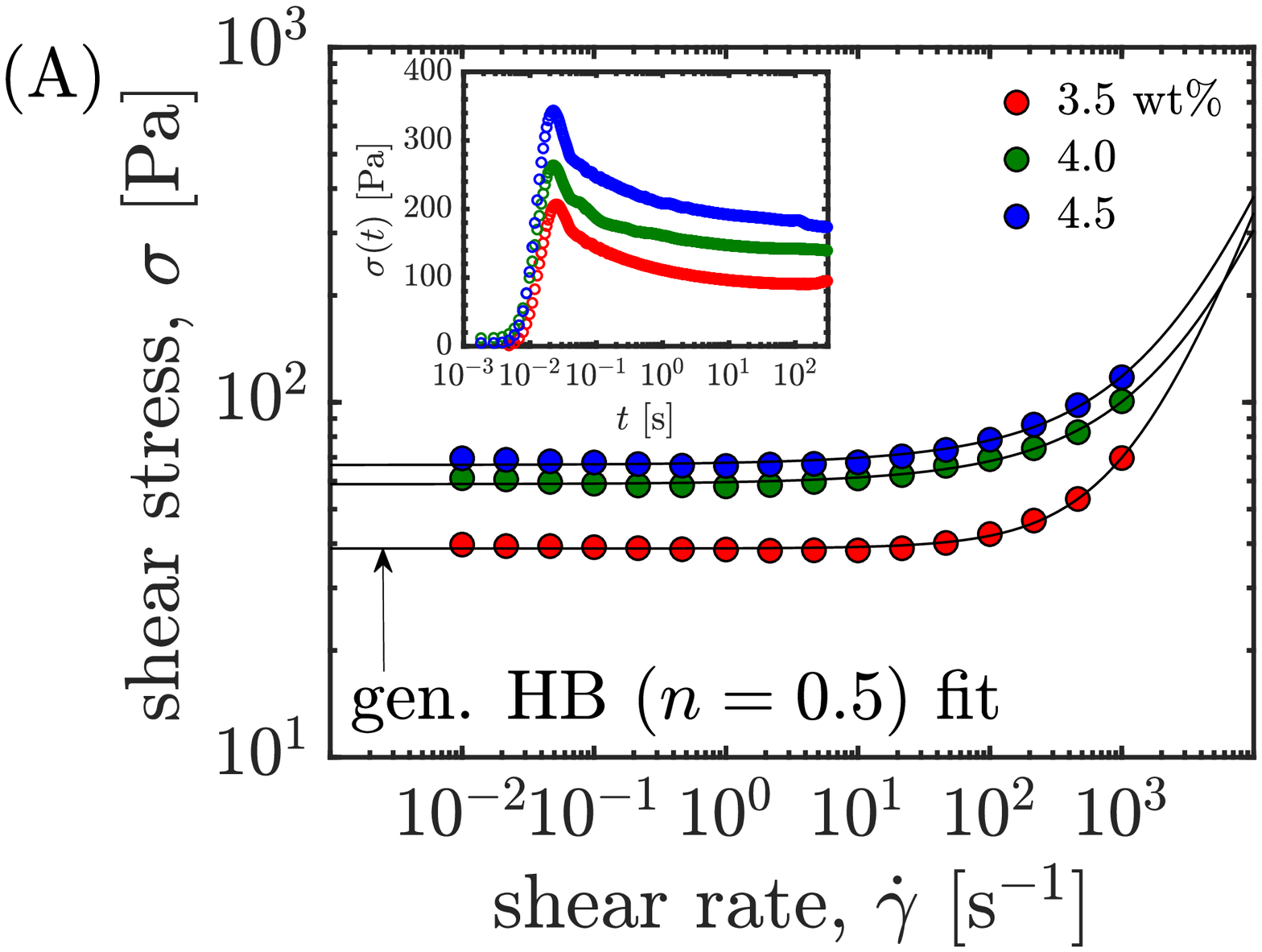}
\end{minipage}
\begin{minipage}{0.49\textwidth}
\centering
    \includegraphics[scale=0.4]{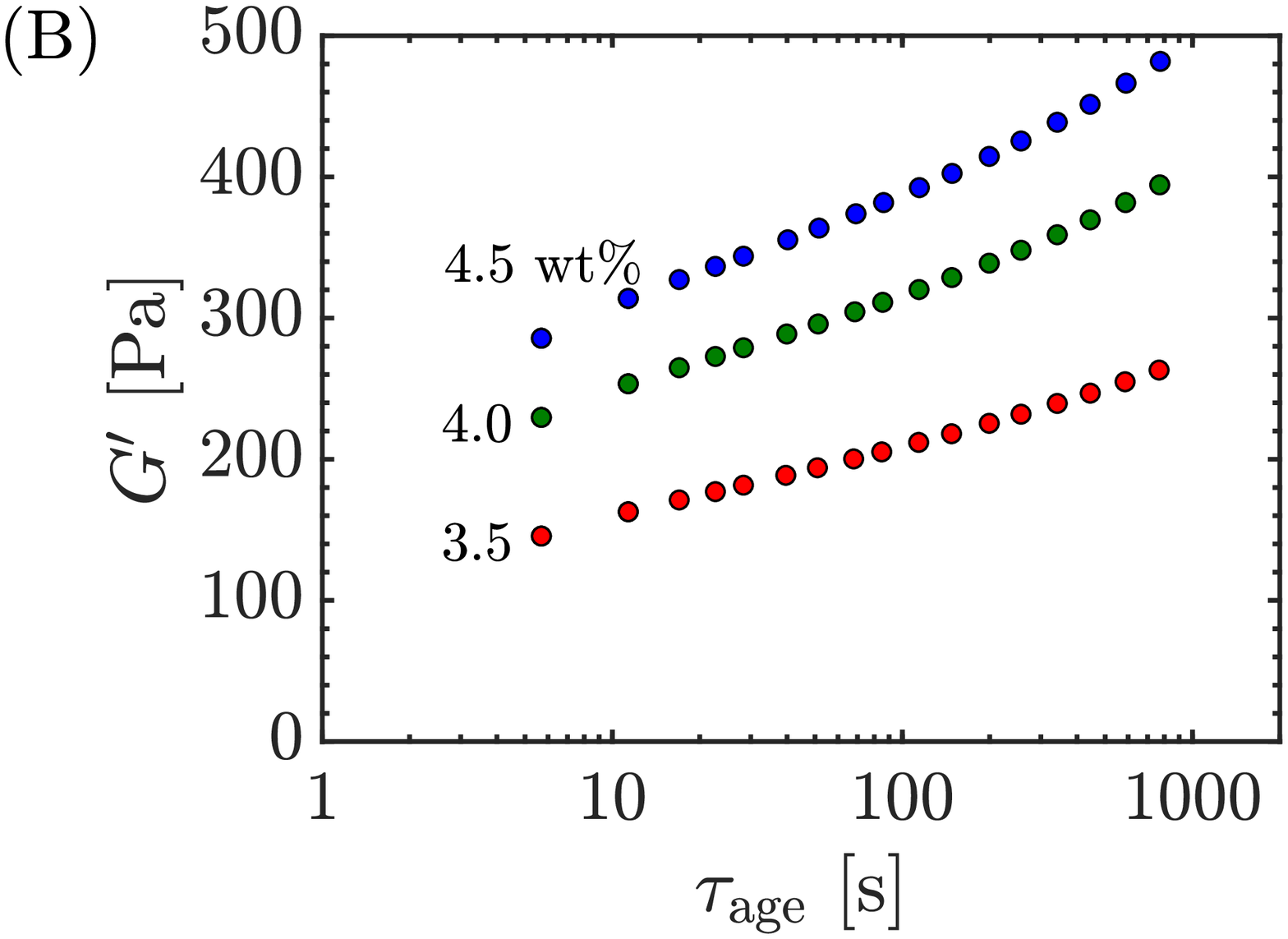}
\end{minipage}
   \caption{\label{fig:laponiteflow}Rheological characterization of Laponite. (A) Steady-state flow curves with fits to the generalized HB model (Eq.~\ref{eq:GenHB}), giving the parameters $\sigma_{\rm y}$, $K$, and $\eta_{\infty}$. (B) Thixotropic stress recovery (aging at rest) of the storage modulus $G^{\prime}$. Tests were conducted within 10~s for ``unaged'' samples, whereas ``aged'' samples were aged for 10~min. The inset in (A) shows the stress breakdown during a startup test for aged samples at each concentration, at the highest rate of $\dot{\gamma}_{\rm f} = 100~\rm{s}^{-1}$ (for the complete data, see Sec.~3, SI). Adapted from \cite{SSRHE_JFM2020} with permission.}
\end{figure}

Rheological characterization of Laponite samples in simple shear (steady shear as well as thixotropic breakdown and recovery) was done with rotational rheometers with parallel plate geometries, as previously described in \cite{SSRHE_JFM2020}. Data for each is shown in Fig.~\ref{fig:laponiteflow}. Since a suspension of Laponite in water is noticeably thixotropic (Fig.~\ref{fig:laponiteflow} (B)), we carefully conduct drop impact tests before the structure recovers or ages significantly to truly test ``unaged'' samples, and hence neglect the effect of thixotropic aging. We expect aging to play a major role in drop impact dynamics by changing the material properties. The effect of aging is studied by testing samples that have been allowed to sit for 10~min. This aging time was chosen because it allows for significant aging, but is short enough for our large number of experiments to be conducted within a feasible time frame. Thixotropic effects can wreck scaling relations, as acknowledged by \cite{LuuForterre2009}. In Fig.~\ref{fig:laponiteflow} (B), the recovery of the structure of Laponite is indicated by monitoring the growth of the linear viscoelastic storage modulus, $G^{\prime}(\omega; \tau_{\text{age}})$ (at a fixed frequency, $\omega = 10~\text{rad~s}^{-1}$, and oscillation strain amplitude, $\gamma_0 = 1\%$), over time of aging, $\tau_{\text{age}}$. The aging time, $\tau_{\text{age}}$, is measured from the point of cessation of a pre-shear applied to the samples at $\dot{\gamma} = 100~\text{s}^{-1}$ for 100~s.

Thixotropic breakdown data was obtained in start-up of steady shear tests, where the samples were allowed to rest for a set period of time ($\sim10$~s for unaged, as a control case, and $\sim600$~s for aged), which was followed by a step increase in shear rate to a larger value, $\dot{\gamma}_{\rm f}$. The inset in Fig.~\ref{fig:laponiteflow} shows the stress during a startup test for aged samples at each concentration, with $\dot{\gamma}_{\rm f} = 100~\rm{s}^{-1}$. Within $\sim100~\rm{ms}$ (relevant to drop impact duration), significant breakdown occurs. Breakdown continues, though more slowly, even beyond 10~s. Stress recovery tests in step shear were also performed, where samples were sheared at a high rate before stepping down to a much lower shear rate. The complete data for these tests are shown in the SI (Sec.~3 and 4 respectively).

Steady-state rheological data, as in Fig.~\ref{fig:laponiteflow} (A), were fit to the generalized Herschel-Bulkley viscoplastic model \cite{SouzaMendes2004}, which is given in one-dimensional steady shear as

\begin{align}\label{eq:GenHB}
\sigma = \sigma_{\rm y} + K \dot{\gamma}^n + \eta_{\infty} \dot{\gamma},
\end{align}
where $\sigma$ is the shear stress. The model parameters $\sigma_{\rm y}$, $K$, $n$, and $\eta_{\infty}$ are the yield stress, consistency index, flow index, and the infinite-shear viscosity, respectively. We enforce $n=0.5$ for all cases when fitting the model to data. This assumption is in agreement with the literature on soft-particle glasses and particulate gels where this Herschel–Bulkley scaling exponent, $n$, is found to be close to 0.5 \cite{Cloitre2003}.

\section{Results\label{sec:results}}

\subsection{Drop impacts and phenomenology\label{subsec:dropimpacts}}
\begin{figure}[!ht]
  \centering
    \includegraphics[scale=0.11]{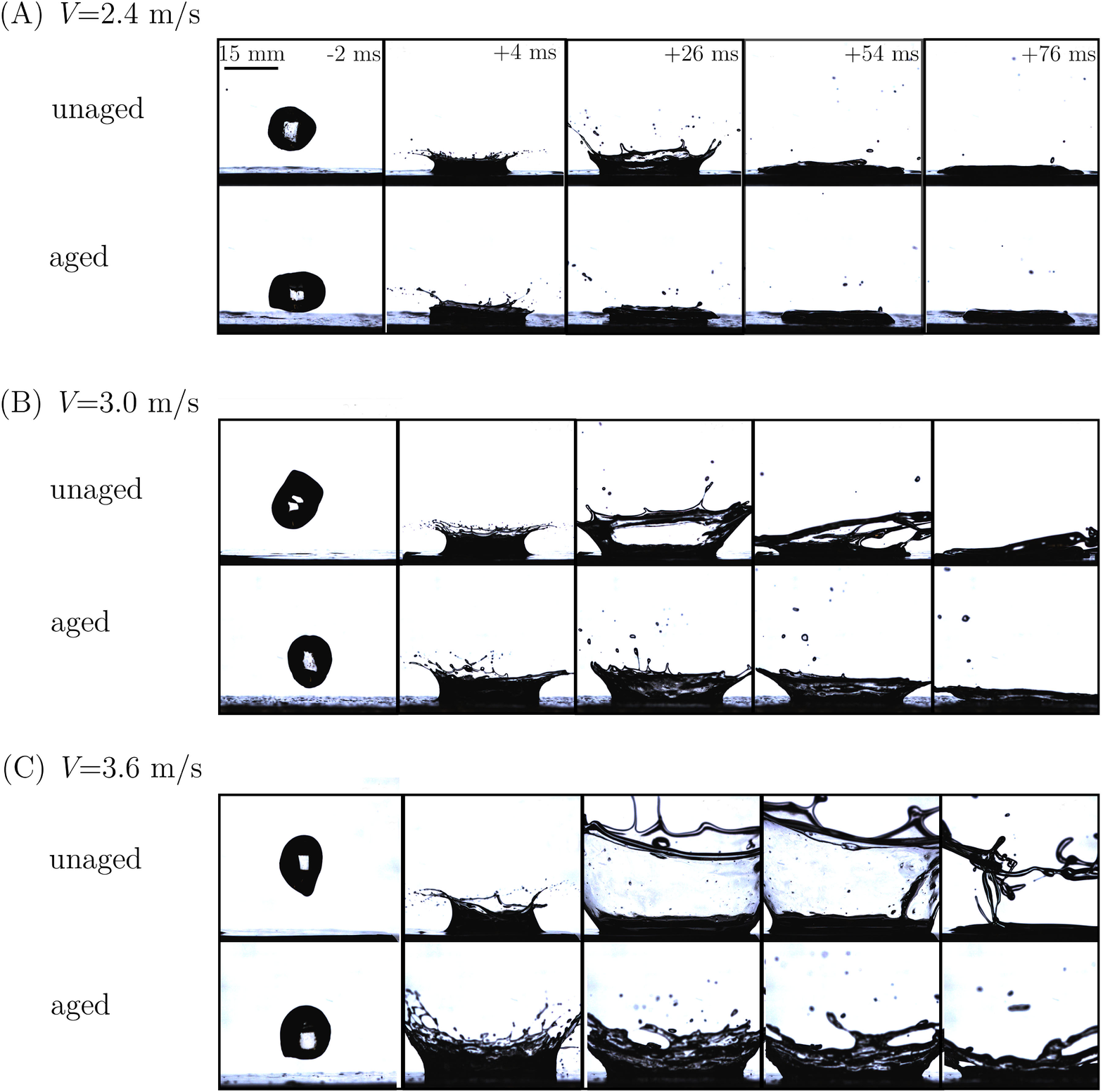}
  \caption{\label{fig:velocityvaried}Effect of aging at varying impact velocities. Drops of $D=15$~mm, 3.5~wt\% Laponite impacting a $t=2.88$~mm thick coating layer of correspondingly same materials at (A) $V=2.4$, (B) $3.0$, and (C) $3.6$~m~s$^{-1}$. For the same experimental conditions, aged samples splash less, i.e.\ less material spatters away from the impact site. Supplementary videos for these drop impact test cases can be found in the online SI.}
\end{figure}

Fig.~\ref{fig:velocityvaried} shows representative impact results from among the 1,008 different conditions tested, shown via the single frames from the videos at key instances of time. It shows a comparison between three different velocities while the other parameters (concentration, $c$, drop diameter, $D$, and coating thickness, $t$) were held constant (we show several additional representative results for the effect of varying each of the four parameters in the SI). The time $\tau=0$ is taken at the instant of impact, so the first tile in each row is shown in negative time relative to the impact frame; time evolves from left to right. As we move from an impact velocity of 2.4 to 3.0 to 3.6~m$\:$s$^{-1}$, the impact event involves more splash and ejection of fluid away from the impact site; this can be detected by either looking at the height of the top of the ejection sheet crown from its base, or at the diameter of the impact crater formed in the coating itself. Also, for each set of two horizontal panels, we compare impact events between unaged and aged samples for the same experimental parameters. We see that the behavior is more ``stick'' than ``splash'' for the aged, and this is evidence of the fact that samples get stiffer or more viscous as they age.

\begin{figure}
  \centering
    \includegraphics[scale=0.45]{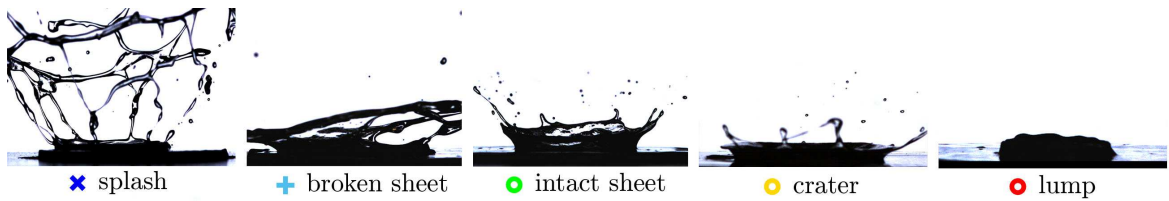}
  \caption{\label{fig:typicalimpactevent}Various impact events observed during experiments, following \cite{SSRHE_JFM2020}. The first two types, splash and broken sheet, are labeled as ``splash'' behaviors, while the last three types, intact sheet, crater, and lump, are called the ``stick'' type behaviors. The impact event types shown here are used as references for judging the type of impact, and the legends used here are consistent with those used in the regime maps that follow. Adapted from \cite{SSRHE_JFM2020} with permission.}
\end{figure}

The type of impact and ejection event is classified as one of five different regimes as shown in Fig.~\ref{fig:typicalimpactevent}, following prior convention in \cite{BCB_PhysFluids2015,SSRHE_JFM2020}. The first two types, called a splash and a broken sheet, are classified as ``splash'' type behaviors, and the last three, called intact sheet, crater, and lump, are the ``stick'' type behaviors. Very little variation is observed between duplicates in terms of the type of resulting impact. For the very few cases where there is a variation, which always occurred near the boundary between two impact types, we designated it to be of the more ``splash'' type of the two. Since there might have been some variability in the time within which the test was conducted, and the Laponite could have aged more than its intended aging state as a result, it is safe to assume the impact would have been more ``splash'' than ``stick'' if the sample had not aged more than intended.

Using this classification, we can determine the impact types for each event in Fig.~\ref{fig:velocityvaried}. In (A), for the unaged sample, we observe an ejection sheet in the frame at +26 ms, which then collapses, but remains intact. This is an intact sheet. For the aged sample, we observe a much smaller ejection sheet, which retracts quickly, only leaving a depression at the impact site. This we call a crater. In (B), for the unaged sample, the ejection sheet is again formed at +26 ms, but this breaks up in the subsequent frame at +54 ms, which then collapses, and hence this is a broken sheet. For the aged sample, the ejection sheet does not break up, and we get an intact sheet. In (C), for the unaged sample, a large ejection sheet is formed compared to the two lower velocities, and this keeps expanding until eventually rupturing into many fragments, as seen in the frame at +76 ms. This is called a ``splash''. The ejection sheet only breaks into a few fragments for the aged sample, and is only a broken sheet. Note that we observe the so-called ``prompt'' splash in almost all impact events. What we choose to broadly call a ``splash'' type behavior is typically referred to as a ``late'' splash in the literature \cite{CossaliMarengo1997}. So we look for a delayed rupture of the ejecta sheet, which happens well beyond the prompt splash that happens immediately upon impact. Also note that the stickier the event, the faster it reaches completion in that the kinetic energy is dissipated more quickly. So, as we move from (A) to (C), the event completion is delayed further. For (C), the ruptured sheet has not even collapsed completely in the final frame shown.

As the samples age, each impact event is less ``splashy,'' i.e.\ less material spatters away from the target surface at the point of impact. The increase in dissipative flow stress is evident. This is where breakdown timescales ($\tau_{\text{thixo},-}$) can influence the results, because depending on how short or long the drop impact event is (characterized by its own experimental timescale, $\tau_{\text{exp}}$) compared to the breakdown timescales, three regimes are possible. Regime $\textcal{1}$ is for very fast breakdown, $\tau_{\text{thixo},-}\ll\tau_{\text{exp}}$, where the aged properties would not matter. The structures will have broken down fast enough for the dissipative stress levels in the fluid to be similar for aged and unaged samples. So, by the time the impact event has proceeded sufficiently to develop characteristic features of the drop and film deformation that are used to classify it, the dissipative forces involved would not be distinguishable between aged and unaged samples. Since we focus on the outcome of the impact event, this regime $\textcal{1}$ would correspond to very similar impact types for aged and unaged samples, even though the impact momentum needed to yield the fluid and initiate deformation would still be higher for aged samples. In such a scenario, steady flow properties would still govern the energy dissipation and consequently the impact outcome, and there would be no need to modify the dimensionless group. Regime $\textcal{2}$ is where $\tau_{\text{thixo},-}\lesssim\tau_{\text{exp}}$, and partial breakdown of the structures would take place before the impact event reaches completion, and this is the most challenging situation to address. Depending on how $\tau_{\text{thixo},-}$ and $\tau_{\text{exp}}$ exactly compare, the effective properties would be in-between the aged and unaged properties at rest. In such a scenario, there may be no simple way to implement effective flow properties into the dimensionless group of Eq.~\ref{eq:IFintro}. Finally, in regime $\textcal{3}$, where $\tau_{\text{thixo},-}\gtrsim\tau_{\text{exp}}$, almost the entire effect of the stiffer, thixotropically aged structures is felt \textit{via} large dissipative forces during the impact event. The difference between aged and unaged samples would be very distinct. In such a case, it may be reasonable to define effective flow properties for aged samples during the comparatively short impact event.

In our experiments, the timescales associated with the drop impact, $\mathcal{O}(10)$~ms, as seen from the time-stamps in Fig.~\ref{fig:velocityvaried}, are either shorter than or comparable to the breakdown timescales of Laponite, $\mathcal{O}(100)$~ms, evident from start-up of shear data for aged samples shown in Fig.~\ref{fig:laponiteflow}(A) at the highest rate of $\dot{\gamma}_{\rm f}$ (for the complete data, see Sec.~3, SI). This suggests $\tau_{\rm{thixo},-}\gtrsim\tau_{\rm exp}$, aand we are in regime $\textcal{3}$. The impact stresses during the drop impact tests were $2000~\rm{Pa}<\frac{1}{2}\rho V^2<18000$~Pa, while the highest stresses during the startup of shear tests were at most 300~Pa. We consider the largest shear rates for estimating some effective value of $\tau_{\text{thixo},-}$, since it is generally expected to decrease with applied shear rate, here approaching a constant of $\approx100$~ms for $\dot{\gamma}_{\rm f}=100$~s$^{-1}$. The structural breakdown upon impact is thus slow compared to the impact timescales, and drop impact should not fully rejuvenate the samples. This may not be the case for other thixotropic fluids if the breakdown timescales become much shorter than drop impact timescales, the case of regimes $\textcal{1}$ and $\textcal{2}$ above.

Whether this increased dissipation is part of the plastic (rate independent) or viscous (rate dependent) rheological response is not yet clear. Projecting the large number of experimental observations onto the dimensionless group of Eq.~\ref{eq:IFintro} will lead toward an answer. We show that it is primarily of plastic nature in Sec.~\ref{sec:agedprops}.

\subsection{Dimensionless numbers and impact regimes\label{subsec:nondimnumbers_regimes}}
We use the classification scheme (Fig.~\ref{fig:typicalimpactevent}) to assign a category to each of the observed impacts, allowing us to make regime maps. The symbols shown in Fig.~\ref{fig:typicalimpactevent} beneath each impact type are used to code the impact regime maps, with each point corresponding to a specific experimental configuration which is represented by the $y-$ and $x-$coordinates (which correspond to the specific values of experimental parameters, $V$, $D$, $t$, and $c$).

Impact regime maps where each point represents the impact type observed (one of the five aforementioned impact types) are used to collapse the five-dimensional space into a two-dimensional plot. To do this, we need to use a scaling relation or a dimensionless grouping. The proposed dimensionless number from \cite{BCB_PhysFluids2015} takes into account the competing effects of inertial and dissipative flow forces from shear properties, while ignoring other effects due to extensibility, chemical structure and morphology differences, and surface tension in comparison to these. The values of the Bond number ($\rm{Bo}\sim\mathcal{O}(100)$, comparing gravity with surface tension), the Weber number ($\rm{We}\sim\mathcal{O}(1000)$, comparing inertia with surface tension) and the capillary number ($\rm{Ca}\sim\mathcal{O}(100)$, comparing viscous forces with surface tension) are very large for our test conditions, so surface tension can be neglected compared to viscous and inertial effects. We will test if the group in Eq.~\ref{eq:IFintro}: (i) is capable of distinguishing impact regimes, (ii) provides a constant critical value of $\text{IF}(D/t) \approx C$, (iii) gives comparable critical $C$ values across different concentrations, and most importantly, (iv) how the critical $C$ values compare between aged and unaged samples. With reference to Eq.~\ref{eq:IFintro}, we hypothesize a critical value of this ratio, $C$, above which inertial forces dominate and below which dissipative flow forces govern the behavior. Thixotropy does not explicitly appear, since $\sigma_{\rm y}$, $K$, and $\eta_{\infty}$ are obtained from steady flow data, i.e.\ fully rejuvenated (unaged) conditions. Consequently, we test the hypothesis that a constant $C$ defines the regime boundary between stick and splash impact types, i.e.\ a more ``splash'' type impact occurs if

\begin{align}\label{eq:splashC}
\text{IF}\left( \frac{D}{t} \right) \gtrsim C.
\end{align}

A detailed explanation of the rationale behind this dimensionless group, including the limitations and caveats, can be found in \cite{BCB_PhysFluids2015,SSRHE_JFM2020}. Additionally, even though we use the same group in Eq.~\ref{eq:IFintro} for both aged and unaged samples, it is not yet clear how the dimensionless group should be modified to incorporate thixotropic effects. This will be addressed in Sec.~\ref{sec:agedprops}, where we show that. We shall show that additional modifications need to be made.

\subsection{Regime maps for Laponite\label{subsec:regimemaps_uncorr}}
\begin{figure}
  \begin{minipage}[!ht]{0.32\textwidth}
    \centering
     \includegraphics[scale=0.30,trim={0 0 0 0},clip]{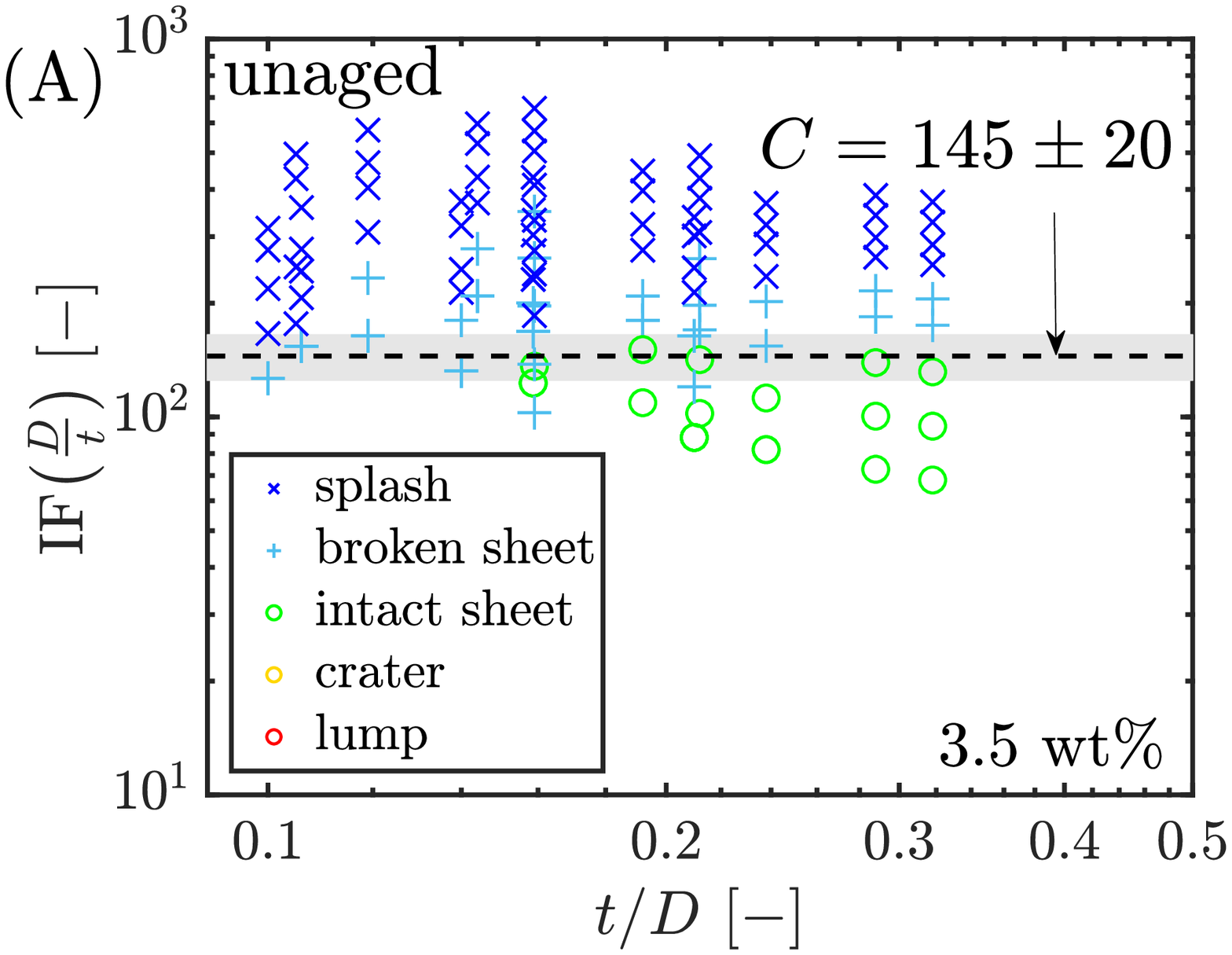}
  \end{minipage}
  \begin{minipage}[!ht]{0.32\textwidth}
    \centering
     \includegraphics[scale=0.30,trim={0 0 0 0},clip]{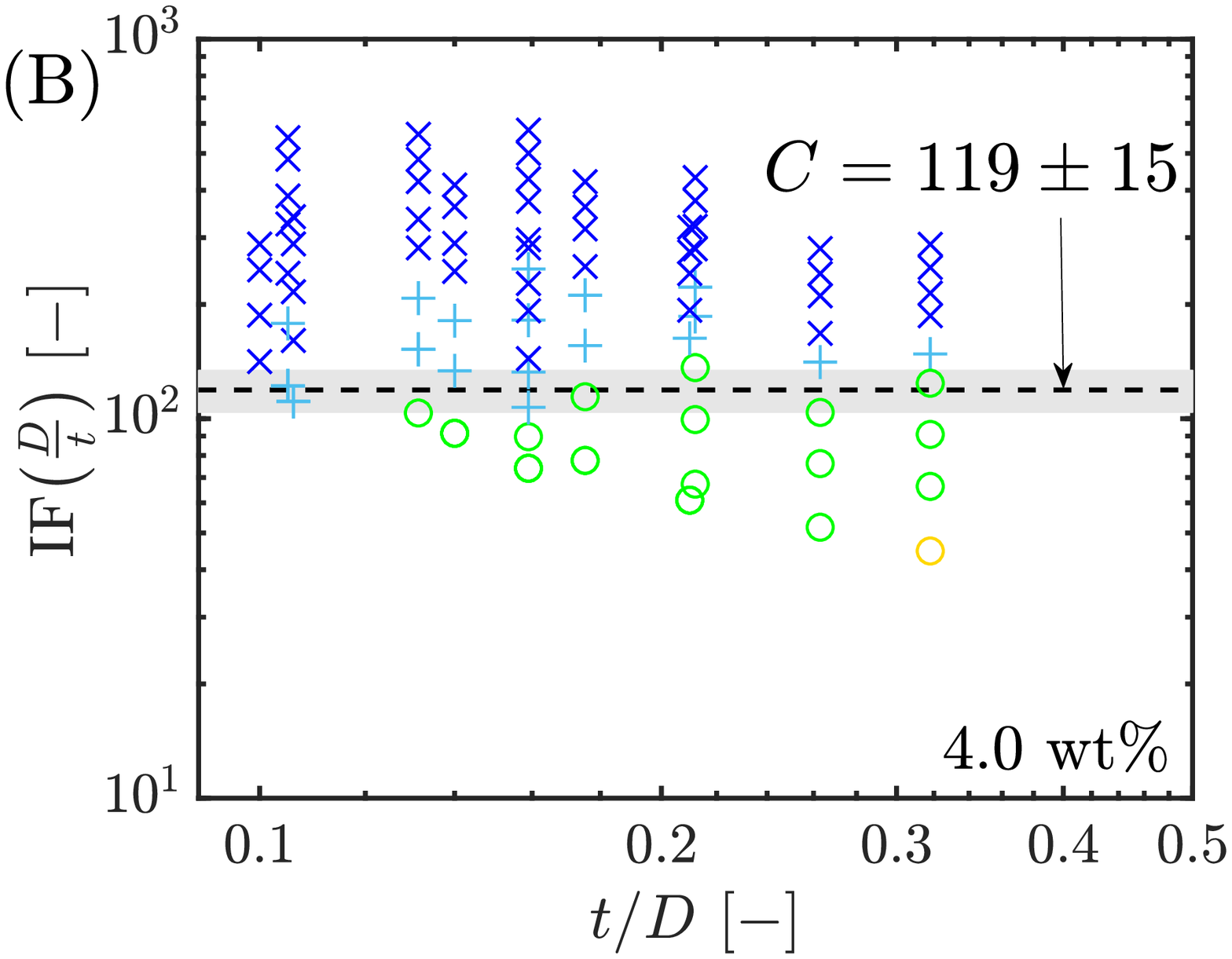}
  \end{minipage}
  \begin{minipage}[!ht]{0.32\textwidth}
    \centering
     \includegraphics[scale=0.30,trim={0 0 0 0},clip]{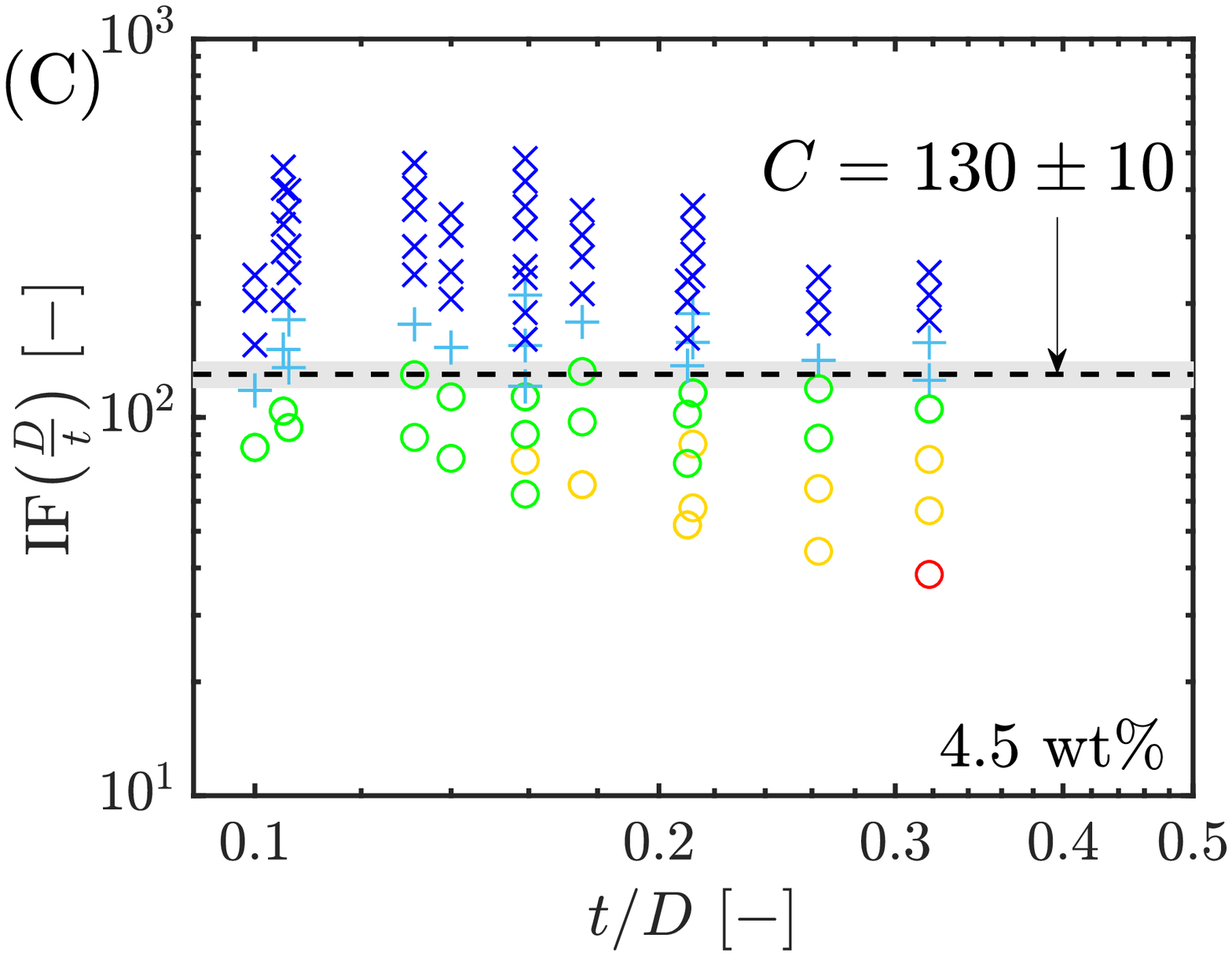}
  \end{minipage}
 \caption{Unaged Laponite regime maps. The steady state properties used are listed in Table~\ref{table:params_unaged}. Adapted from \cite{SSRHE_JFM2020} with permission.\label{fig:regimemaps_unaged}}
\end{figure}

\begin{figure}
  \begin{minipage}[!ht]{0.32\textwidth}
    \centering
     \includegraphics[scale=0.30,trim={0 0 0 0},clip]{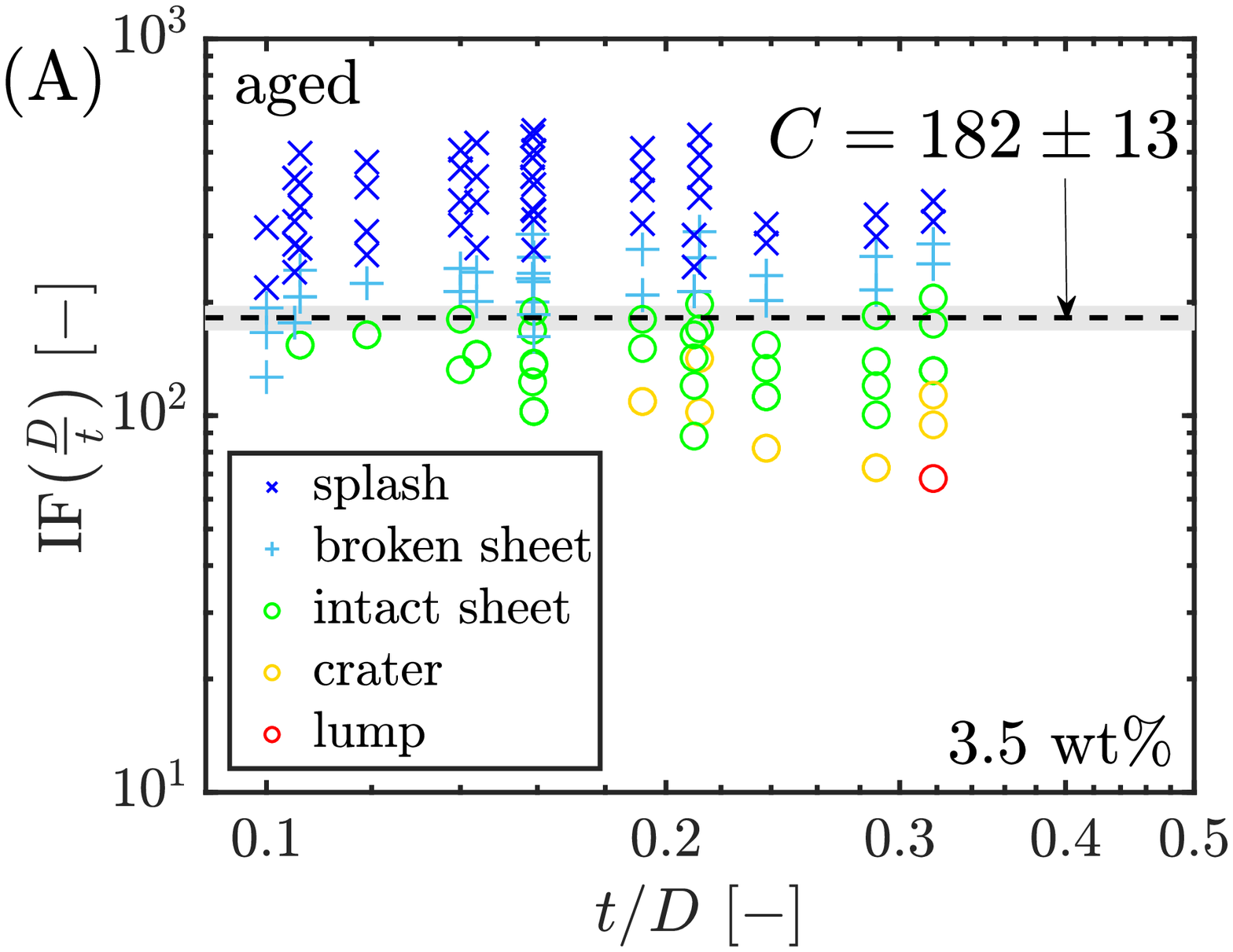}
  \end{minipage}
  \begin{minipage}[!ht]{0.32\textwidth}
    \centering
     \includegraphics[scale=0.30,trim={0 0 0 0},clip]{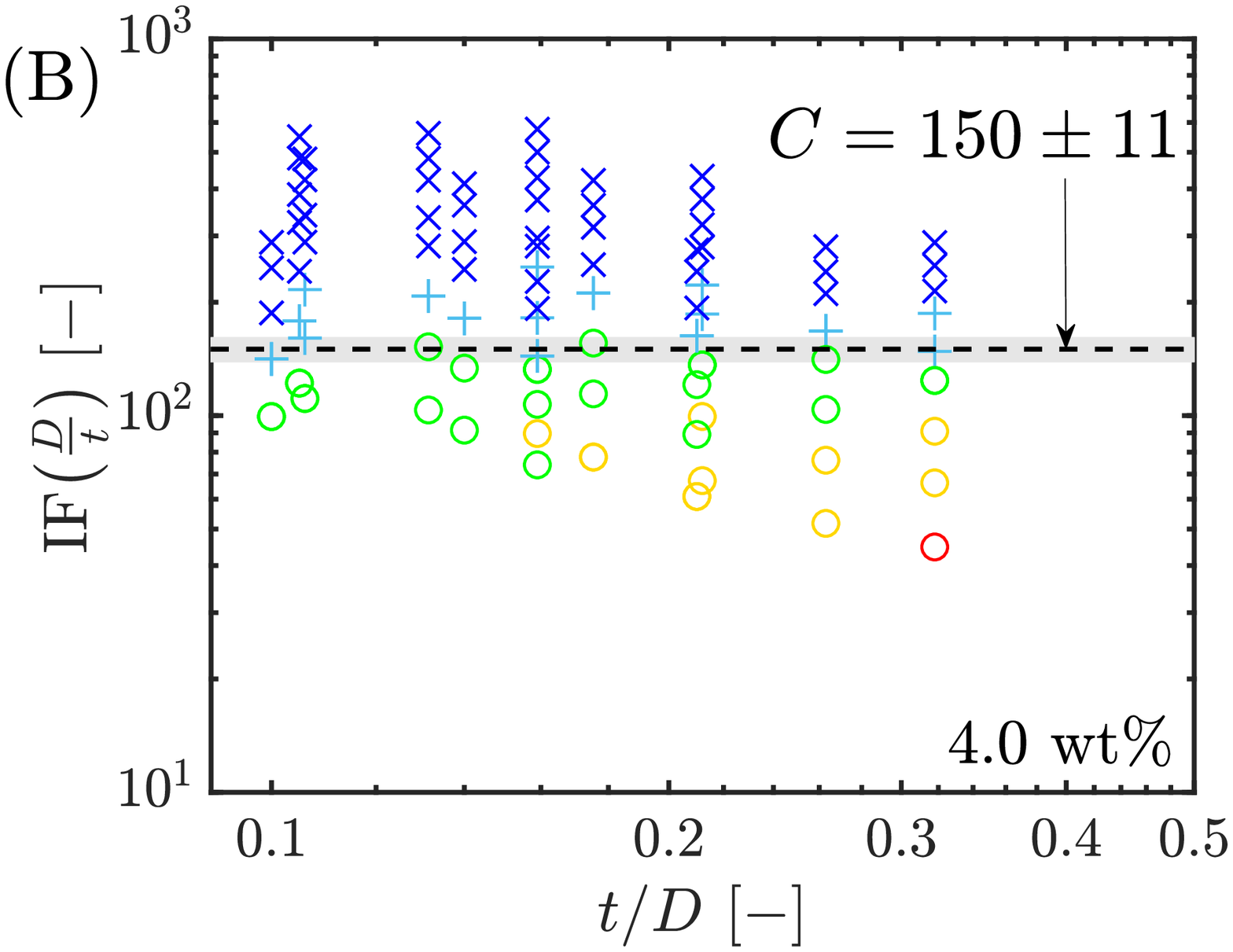}
  \end{minipage}
  \begin{minipage}[!ht]{0.32\textwidth}
    \centering
     \includegraphics[scale=0.30,trim={0 0 0 0},clip]{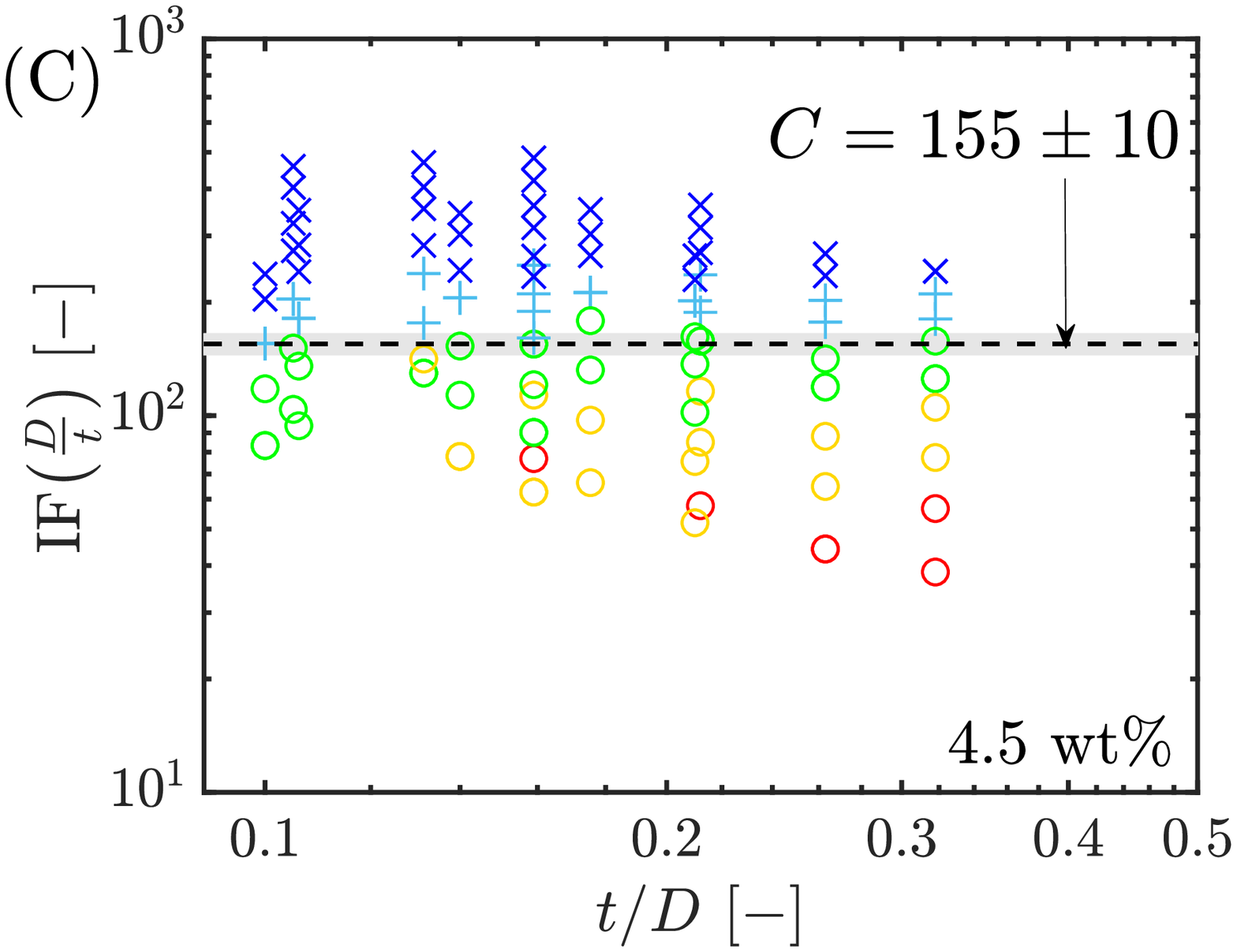}
  \end{minipage}
 \caption{Aged Laponite regime maps. The steady state properties used to plot these are the same as those for the unaged, as listed in Table~\ref{table:params_unaged}.\label{fig:regimemaps_aged_uncorr}}
\end{figure}

The results of the drop impact tests have been grouped by concentration ($c$) of Laponite, in two clusters: unaged and aged samples. Figs.~\ref{fig:regimemaps_unaged}, \ref{fig:regimemaps_aged_uncorr} show the results, plotted using the schemes and conventions mentioned earlier, with the effect of various experimental parameters collapsed into one group, IF$(D/t)$.

For all the concentrations of unaged Laponite, we see that the different impact types are effectively separated via this non-dimensionalization. There is very little overlap between different symbols representing the different impact event types. This is a very encouraging result in itself, and shows that this dimensionless group is able to capture the leading order physics, and also supports the claim to classify the impact events in the manner that we have done here. We also see that a boundary exists between the ``splashy'' and ``sticky'' regimes, which is chosen to be that between \emph{broken sheet} and \emph{intact sheet}. This boundary is of almost zero slope, suggesting a constant critical value of $\text{IF}(D/t)=C$ for a transition from stick to splash. The value of $C$ is determined, following our previously described method \cite{SSRHE_JFM2020}, by fitting a line with zero slope to the set of data points for the broken sheet with the lowest values of $\text{IF}(D/t)$, and those for the intact sheet with the highest values of $\text{IF}(D/t)$. This fits the regime boundary between these two impact types, and uncertainty on the value of $C$ is from the confidence intervals of the fit. More details can be found in our previous work \cite{SSRHE_JFM2020}. Changing any one or a combination of the four parameters may make an event transition from stick to splash, provided the critical value of $C$ is crossed during the change. For example, if film thickness $t$ is increasing during spray coating, the value of $\text{IF} (D/t)$ will decrease and the impact type can change type from splash to stick.

The success of $\text{IF}(D/t)=C$ is strengthened by our finding of a similar value of $C$ across different concentrations of unaged  Laponite, $C \approx 131$ on average \cite{SSRHE_JFM2020}. The variation is non-monotonic with Laponite concentration and the uncertainty ranges overlap for each formulation. The mean values of $C$ for each concentration are within 10\% of each other.

The group is only partially effective for the aged samples, as per the criteria mentioned in Sec.~\ref{sec:introduction}. The boundary is still a line of more or less constant slope. The impact types are also separated quite well. But the values of $C$ are more disparate across all three concentrations when compared to the differences in unaged samples; the difference in mean values is as high as 20\%. More importantly, they do not overlap even within the fit uncertainty values. Additionally, for a given concentration, the $C$ values do not compare well between aged and unaged samples. So criteria (iii) and (iv) are not satisfied by using steady shear flow properties for plotting maps for the aged. This is not surprising. From Fig.~\ref{fig:velocityvaried}, it is clear that as a sample ages, its properties change in such a way that the motion-retarding flow forces get stronger, thus resulting in more ``stick'' type outcomes. So if one could estimate the flow properties of aged samples and use these to plot impact regime maps, perhaps all the criteria might be satisfied. Using this as a motivation, we proceed to indirectly estimate the flow properties of aged samples, i.e.\ updating the denominator of IF. These indirect measures can be used to define updated dimensionless groups. If these are a sufficiently accurate reflection of aged flow properties, they must satisfy all the four criteria outlined in Sec.~\ref{sec:introduction} for the validity of the dimensionless group.


\section{Estimating properties of thixotropically aged Laponite\label{sec:agedprops}}

\subsection{Rationale\label{subsec:agedprops_rationale}}
Here, we propose and test possible hypotheses for modifying the effective rheological fluid properties ($\sigma_{\rm y}$, $K$, and $\eta_{\infty}$) to account for thixotropic aging, where $\sigma_{\rm y}$ is the plastic (rate independent) and $K$ and $\eta_{\infty}$ are the viscous (rate dependent) components of the dissipative flow stress. Each of these properties is well-defined for unaged (rejuvenated) samples, and the values are obtained from steady shear curves. We postulate that the effective properties of the \textit{aged} samples modify the denominator of the dimensionless group, generally expressed as

\begin{align}\label{eq:eqn_IF_2}
\text{IF}\left( \frac{D}{t} \right) \sim \frac{\text{inertial forces}\left( \rho, V, D \right)}{\text{dissipative flow forces}\left( \sigma_{\rm y}, K, \eta_{\infty}, V, t; \tau_{\text{age}} \right)},
\end{align}
where the flow forces are now also a function of the aging time, $\tau_{\text{age}}$. At small values of $\tau_{\text{age}}$, the sample is relatively unaged, and using steady shear flow properties is justified; this is evident from the fact that the regime maps satisfy criteria (i)-(iii) at the end of Sec.~\ref{sec:introduction}. To plot maps corresponding to aged samples, one approach is to simply allow $\sigma_{\rm y}$, $K$, and $\eta_{\infty}$ to each be a function of $\tau_{\text{age}}$. Assuming the dissipative flow stresses retain their form, the dimensionless group can be rewritten as

\begin{align}\label{eq:eqn_IF_3}
\text{IF}\left( \frac{D}{t} \right) \equiv \frac{\rho V^2 D}{\left[ \sigma_{\rm y}(\tau_{\text{age}}) + K(\tau_{\text{age}})(V/t)^{0.5} + \eta_{\infty}(\tau_{\text{age}}) V/t \right]t},
\end{align}
where $\sigma_{\rm y}$, $K$, and $\eta_{\infty}$ are functions of $\tau_{\rm age}$. Other suitable forms of expressing the rheological properties accounting for aging might exist, but we hypothesize that the original expression used for unaged samples is retained. One would naturally assume that, due to aging, either one or more of these properties increase with $\tau_{\text{age}}$, and that would explain why the fluids become ``stickier''. This rationale is substantiated by the fact that the critical $C$ obtained for aged samples is greater than those for the unaged, since instead of using the \textit{aged} values of the flow properties, we are using the smaller, \textit{unaged} properties to plot the maps in Fig.~\ref{fig:regimemaps_aged_uncorr}. Using the more accurate aged properties should increase the dissipative contribution in the denominator in Eq.~\ref{eq:eqn_IF_3}, and the regime boundaries should move lower, closer to those for the unaged samples. In the same spirit, we can assume that each of $\sigma_{\rm y}$, $K$, and $\eta_{\infty}$ possibly change (typically increase) due to aging. Let us define a ratio, $\varphi_{\psi}$, which is the ratio of aged to unaged values of a fluid property, $\psi$, such that

\begin{align}\label{eq:varphi1}
\varphi_{\psi} \equiv \frac{\psi^{\text{aged}}}{\psi^{\text{unaged}}} \equiv \frac{\hat{\psi}}{\psi},
\end{align}
where, for compactness of notation, aged properties are denoted with a \textit{hat} $(\hat{\psi})$, while unaged properties without any $(\psi)$, e.g., for steady flow properties. This form does not assume whether the properties increase or decrease by aging. The aged yield stress, consistency index, and infinite shear viscosity can accordingly be written as

\begin{subequations}
\label{eq:aged_varphi_all}
\begin{align}
\hat{\sigma}_{\text{y}} &= \varphi_{\sigma_{\rm y}} \sigma_{\rm y}, \label{eq:aged_varphi_ys}\\
\hat{K} &= \varphi_{K} K, \label{eq:aged_varphi_K}\\
\hat{\eta}_{\infty} &= \varphi_{\eta_{\infty}} \eta_{\infty}, \label{eq:aged_varphi_eta}
\end{align}
\end{subequations}
respectively. The dimensionless group can now be modified to include the effect of aging as

\begin{align}\label{eq:eqn_IF_4}
\text{IF}\left( \frac{D}{t} \right) \equiv \frac{\rho V^2 D}{\left[ \varphi_{\sigma_{\rm y}} \sigma_{\rm y} + \varphi_{K}K(V/t)^{0.5} + \varphi_{\eta_{\infty}}\eta_{\infty} V/t \right]t}.
\end{align}
The challenge lies in determining the factors $\varphi_{\sigma_{\rm y}}$, $\varphi_{K}$, and $\varphi_{\eta_{\infty}}$.

\begin{figure}[tb]
  \begin{minipage}[!ht]{0.32\textwidth}
    \centering
        \includegraphics[scale=0.28,trim={0 0 0 0},clip]{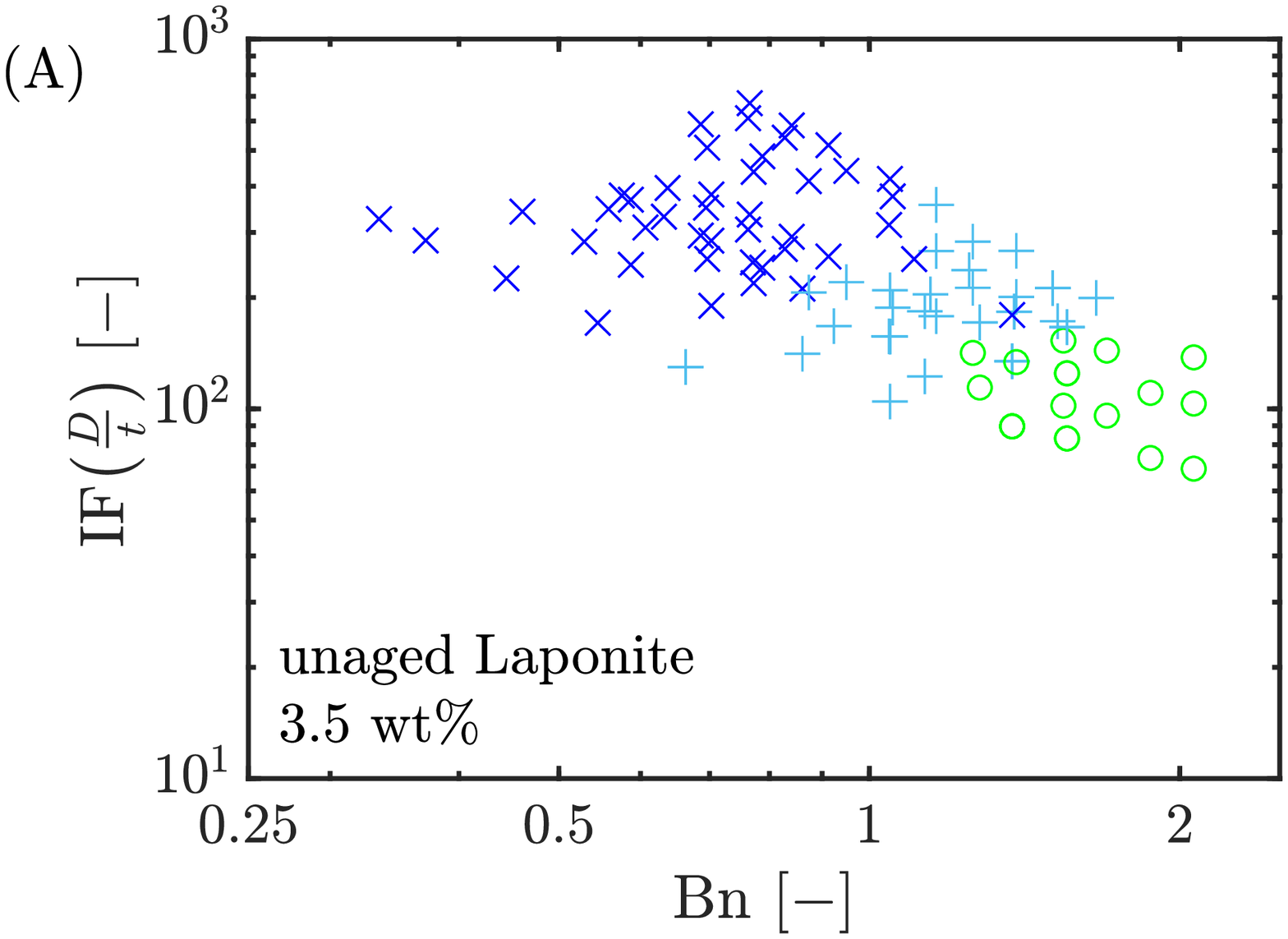}
  \end{minipage}
  \begin{minipage}[!ht]{0.32\textwidth}
    \centering
     \includegraphics[scale=0.28,trim={0 0 0 0},clip]{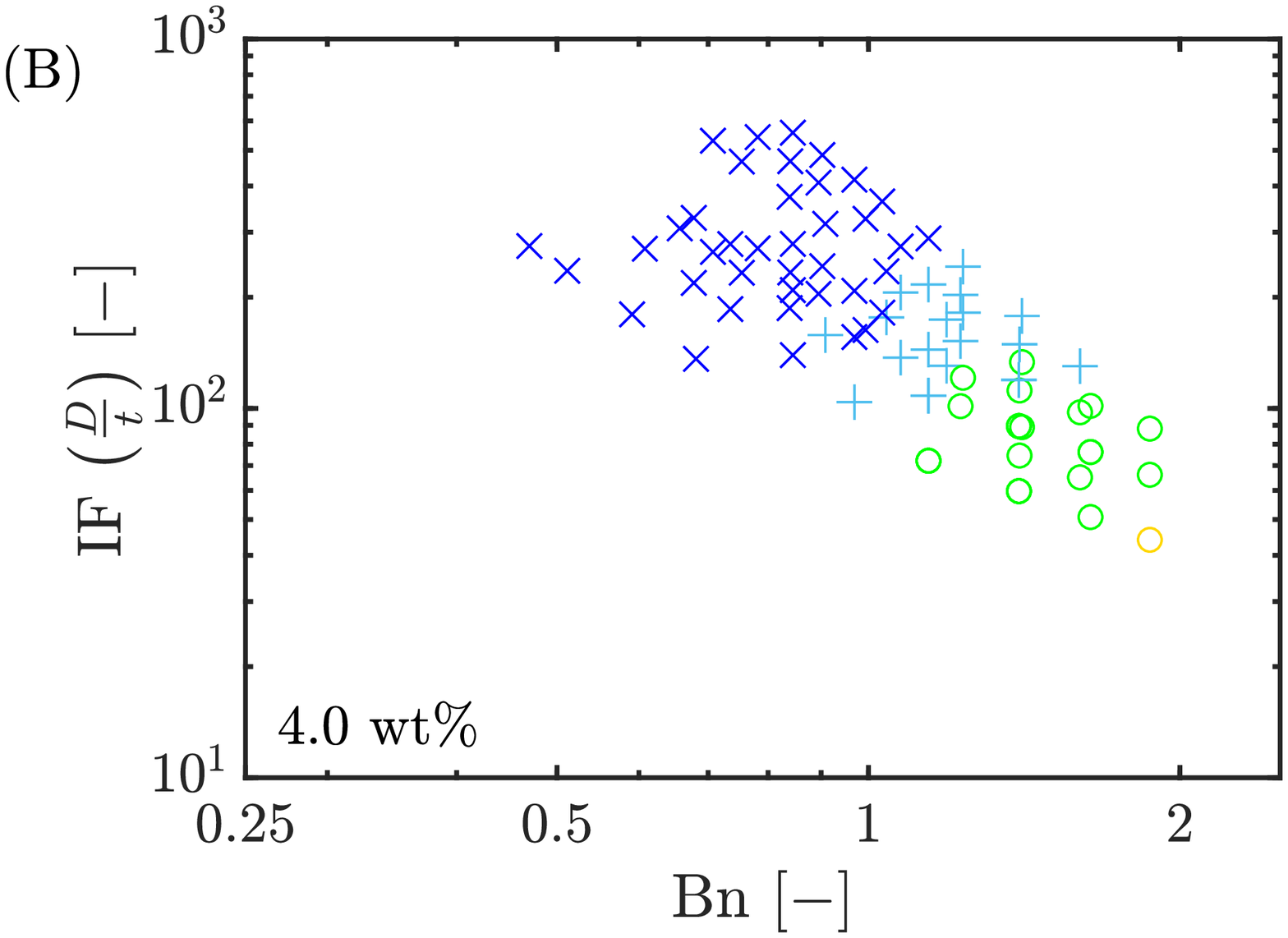}
  \end{minipage}
  \begin{minipage}[!ht]{0.32\textwidth}
    \centering
     \includegraphics[scale=0.28,trim={0 0 0 0},clip]{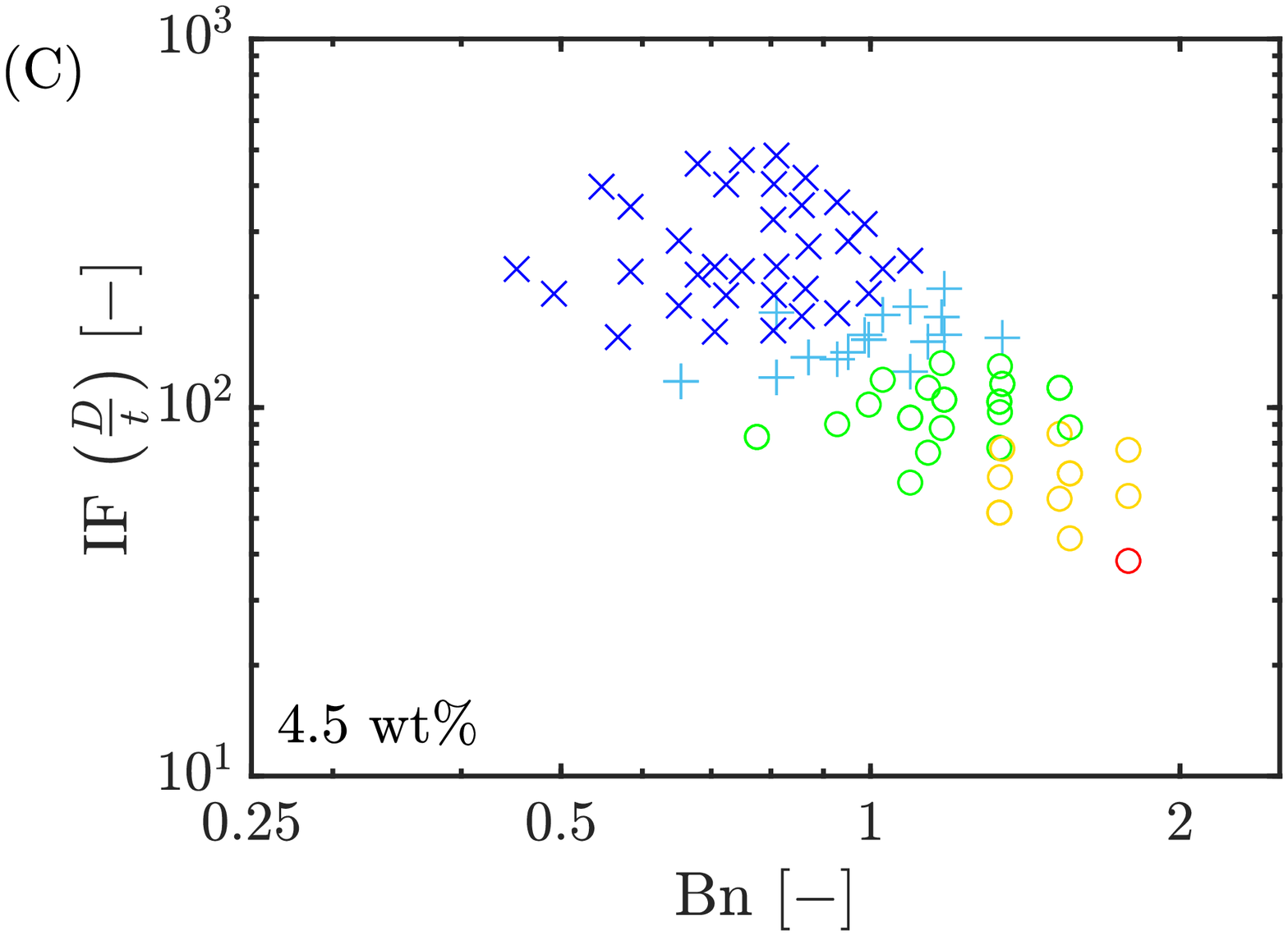}
  \end{minipage}
 \caption{Plots of Bingham number, $\text{Bn}$, defined based on the scale of shear stress for each drop impact event, $\dot{\gamma} \sim V/t$, vs. IF$(D/t)$, for each concentration of (aged) Laponite, plotted using steady-state shear properties. This helps us gauge the dominant component of the flow stress term (i.e., the denominator of IF$(D/t)$). We see that for most cases, especially for the regime boundary of interest, the ratio of yield stress to viscous stress is always $0.1-1$, with the nominal values of $\text{Bn} \approx 0.3 \sim 0.6$.\label{fig:agedregimemapsuncorr_IFvBingham}}
\end{figure}

The relative importance of plastic (rate independent) ($\sigma_{\rm y}$) versus viscous (rate dependent) ($K(V/t)^{0.5} + \eta_{\infty}V/t$) contributions is captured by the Bingham number \cite{ThompsonSoares2016}, defined as

\begin{align}\label{eq:Bn_def}
    \text{Bn} \sim \frac{\text{plastic stress}}{\text{viscous flow stress}} \equiv \frac{\sigma_{\rm y}}{K(V/t)^{0.5} + \eta_{\infty}V/t}.
\end{align}
One can also use the Plastic number, $\text{Pl}=\frac{\text{Bn}}{1 + \text{Bn}}$, which compares the plastic stress to the total shear stress \cite{ThompsonSoares2016,KimFreundEwoldt_JNNFM2019}. We prefer using Bn because this gives a more direct comparison between the two different contributions. From Fig.~\ref{fig:agedregimemapsuncorr_IFvBingham}, we see that the Bingham number is typically smaller than 1, and ranges between 0.3 and 0.6 for the regime boundary chosen to obtain $C$. So, any change to the viscous terms may affect the value of IF$(D/t)$ more than changes in the yield stress, but the changes from modifying either parameter may be significant. So we must look for methods that change (or increase, in our case) the values of $\sigma_{\rm y}$, $K$, and $\eta_{\infty}$ to estimate the flow properties of aged samples, and shift the regime maps significantly. Rheological properties obtained from \emph{steady flow} data, $\sigma_{\rm y}$ $K$, and $\eta_{\infty}$, cannot be defined for aged materials. So, we look for indirect methods of estimating these properties, obtaining expressions for $\varphi_{\sigma_{\rm y}}$, $\varphi_{K}$, and $\varphi_{\eta_{\infty}}$.

\subsection{Hypothesis: change only the yield stress based on aging storage modulus\label{subsec:agedprops_hypothesis_onlyYS}}

We tested multiple hypotheses for finding the effective $\sigma_{\rm y}$, $K$, and $\eta_{\infty}$ for aged samples, and found the strongest evidence for changing only the plastic component $\sigma_{\rm y}$ (and not the viscous $K$ or $\eta_{\infty}$ components) and basing this increase in $\sigma_{\rm y}$ on the observed $G^{\prime}\left(\tau_{\rm age}\right)$, Fig.~\ref{fig:laponiteflow}(B). We call this ``Hypothesis~1,'' to distinguish from the others. This is embodied by the shift factor

\begin{subequations}
\label{eq:hypothesis_onlyYS}
\begin{align}
\varphi_{\sigma_{\rm y}} &= \varphi_{G^{\prime}} \varphi_{\gamma_{\text{y}}} \approx \varphi_{G^{\prime}},   \label{eq:hypothesis_onlyYS_1}\\
\varphi_{K} &= \varphi_{\eta_{\infty}} = 1  \label{eq:hypothesis_onlyYS_2}.
\end{align}
\end{subequations}
Other hypotheses, such as changing each of $\sigma_{\rm y}$, $K$, and $\eta_{\infty}$ (Hypothesis~2), or using static yield-stresses obtained from startup of steady shear tests (Hypothesis~3), are explained and tested in the SI (sections~2 and 3).

Eq.~\ref{eq:hypothesis_onlyYS_1} maps $\sigma_{\rm y}$ to linear elastic $G^{\prime}$, which for our samples is a weak function of the oscillation frequency $\omega$, as has been reported in the literature for gels and colloidal suspensions. Eq.~\ref{eq:hypothesis_onlyYS_2} assumes that $K$ and $\eta_{\infty}$ remain unchanged in spite of aging. But one might expect all flow properties to increase with structure build-up, whether plastic or rate dependent. Changing only $\sigma_{\rm y}$ while keeping $K$ and $\eta_{\infty}$ unchanged may seem counter-intuitive, but the rationale for this hypothesis is that $\sigma_{\rm y}$ may break down sufficiently slowly (following the argument presented in Sec.~\ref{subsec:dropimpacts}, Fig.~\ref{fig:laponiteflow}(A), inset) to dominate dissipative effects in aged samples, while high rate effects associated with $K$ and $\eta_{\infty}$ break down faster than drop impact timescales. Structure breakdown is also a very complicated phenomenon, and increasing only $\sigma_{\rm y}$, as a lumped property for all dissipative effects may capture the dominant effect of yielding to initiate flow, since the effective Bingham number, Bn, maybe be larger when considering static rather than dynamic yield stress. Either way, we find that the effect of aging is sufficiently encapsulated within an increased $\sigma_{\rm y}$, and any possible increase in $K$ or $\eta_{\infty}$ due to aging is not significant for the drop impact results here.

\begin{figure}[tb]
\begin{minipage}{0.49\textwidth}
\centering
    \includegraphics[scale=0.4]{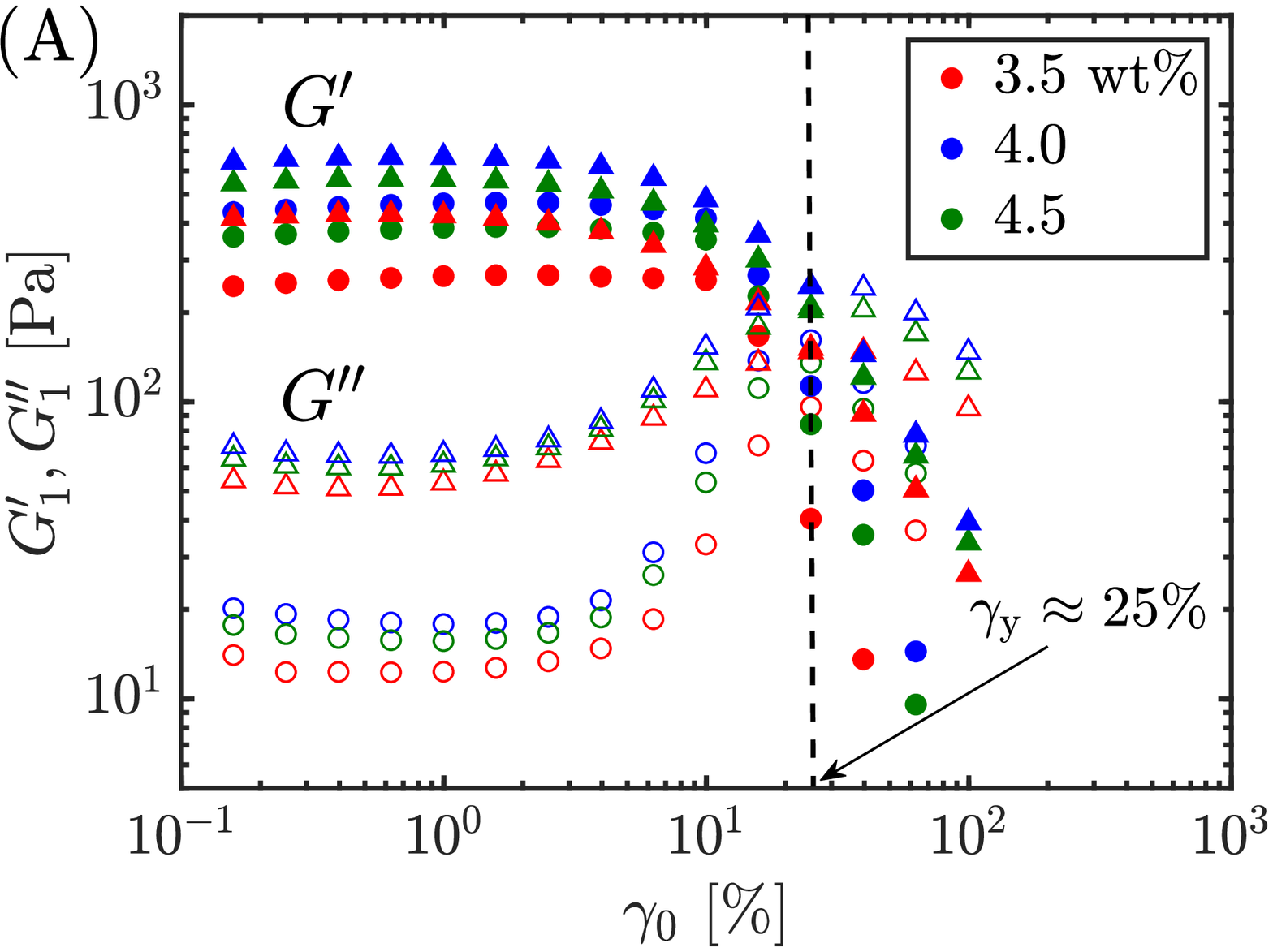}
\end{minipage}
\begin{minipage}{0.49\textwidth}
\centering
    \includegraphics[scale=0.4]{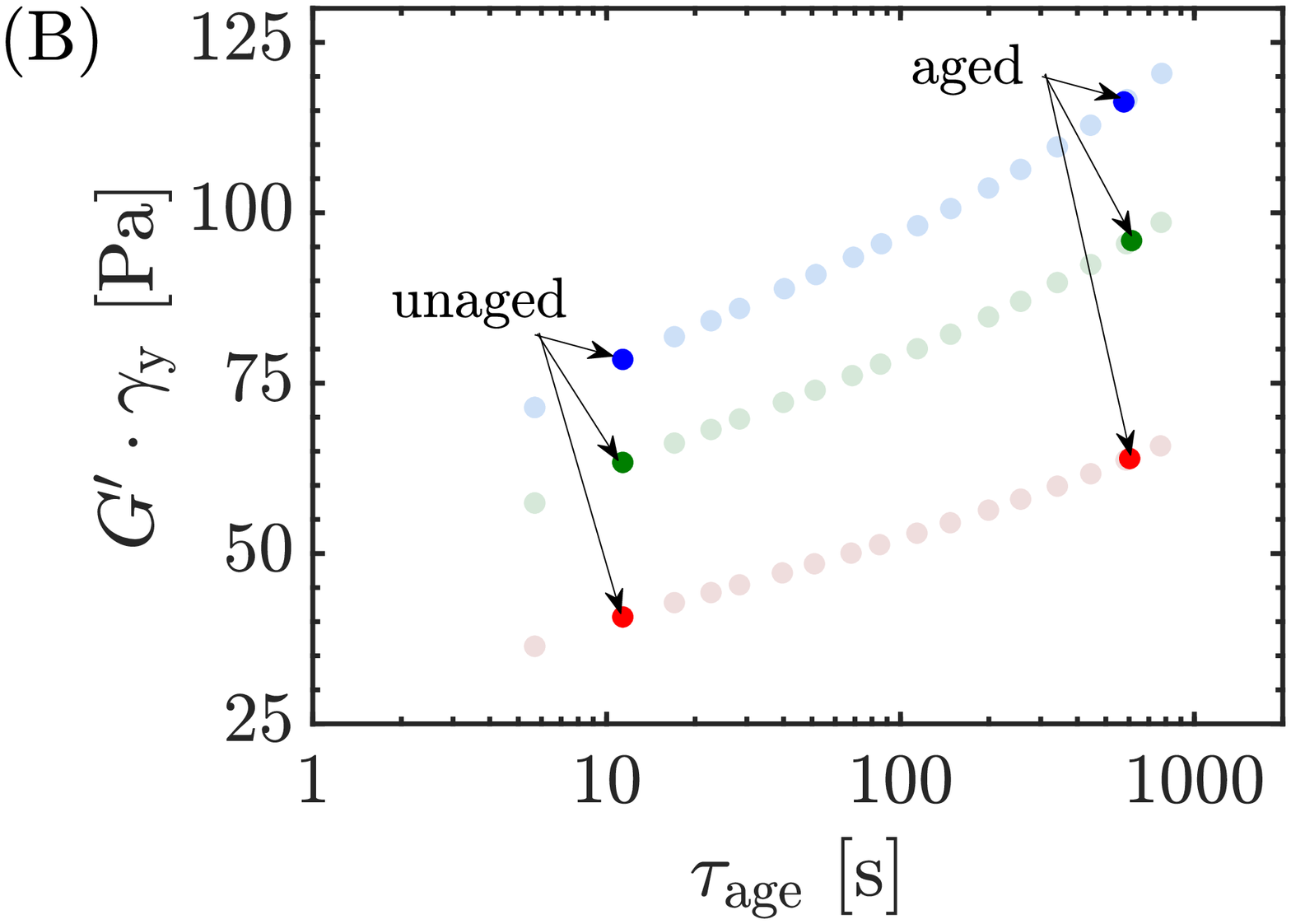}
\end{minipage}
   \caption{(A) Yield strain $\gamma_{\rm y}$ of both unaged ($\bigcirc$) and aged ($\bigtriangleup$) Laponite from oscillatory strain amplitude sweeps at a constant frequency ($\omega=10$~rad~s$^{-1}$, the raw data with parallel plate correction \cite{Wyss_PPCorr2014,Ng_PPCorr2011,KochPhDThesis}). (B) Inferred aging of yield stress, $\sigma_{\rm y} \approx G^{\prime}\gamma_{\rm y}$, using the value $\gamma_{\rm y}\approx25\%$ with $G^{\prime}(\tau_{\rm age})$ data in Fig.~\ref{fig:laponiteflow}(B) \cite{JalaalLohse_JFM2019,Bonn_YS_JNNFM2016}. The values of aged $\sigma_{\rm y}$ obtained for each concentration at an aging time of $\tau_{\text{age}}\approx600\;\text{s}$ are 60, 87, and 99 Pa for 3.5, 4.0, and 4.5 wt\% respectively.\label{fig:GtoYSaging_YS}}
\end{figure}

Yield-stress fluids are often modeled as elastic solids for applied stresses below the yield value \cite{Bonn_YS_JNNFM2016,DullaertMewis_structkinetics2006,Coussot2006}, such that $\sigma \approx G^{\prime}\gamma$. Here, $\sigma$ is the shear stress, $G^{\prime}$ is the linear viscoelastic storage modulus, and $\gamma$ is the shear strain in the sample. Assuming this relation to be valid until the material yields, one can write

\begin{align}\label{eq:GtoYS}
\sigma_{\rm y} \left( \tau_{\text{age}} \right) \approx G^{\prime} \left(\omega; \tau_{\text{age}} \right) \gamma_{\text{y}},
\end{align}
where $\gamma_{\text{y}}$ is the yield strain, or any critical strain measure \citep{DullaertMewis_structkinetics2006,Coussot2006}. The linear storage modulus has weak frequency dependence (see Sec.~5 in the SI). We use a frequency of $\omega=10$~rad~s$^{-1}$ for a large amplitude oscillatory shear (LAOS) amplitude sweep test to measure the linear $G^{\prime}$ and the nonlinear yield strain $\gamma_{\rm y}$. We estimate $\gamma_{\rm y}$ from the crossover of $G^{\prime}$ and $G^{\prime\prime}$, though other methods could be used \cite{DonleyRogers_JNNFM2019}. Such data for Laponite is shown in Fig.~\ref{fig:GtoYSaging_YS}.

Now that we have an expression for the yield-stress as a function of aging time, we can write

\begin{align}\label{eq:varphi_GtoYS1}
\hat{\sigma}_{\text{y}} \approx \hat{G}^{\prime} \hat{\gamma}_{\text{y}} = \left( \varphi_{G^{\prime}} \varphi_{\gamma_{\text{y}}} \right) G^{\prime} \gamma_{\text{y}} = \varphi_{\sigma_{\rm y}} G^{\prime} \gamma_{\text{y}},
\end{align}
where $\varphi_{\sigma_{\rm y}}$ remains to be determined, so we must find expressions for $\varphi_{G^{\prime}}$ and $\varphi_{\gamma_{\text{y}}}$. One way of obtaining $\varphi_{G^{\prime}}$ is to monitor the storage modulus at a fixed frequency and amplitude (in the linear regime) with aging time, as in Fig.~\ref{fig:laponiteflow}(B). These samples were pre-sheared at 100~s$^{-1}$ for 100~s, after which a small oscillatory strain of amplitude $\gamma_0 = 1\%$ at a frequency of $\omega=10$~rad~s$^{-1}$ was immediately applied, and the elastic modulus was monitored with time. As we can see, $G^{\prime}$ increases with time, and is an indicator of aging in the sample \cite{Coussot2006}. We can pick out the values of $G^{\prime}$ at two locations of interest: one at short aging times, corresponding to the unaged sample, and one at longer aging times, for the aged state. In our specific case, the unaged and aged states were chosen to be those at $\tau_{\text{age}} = 10$~s and 600~s, corresponding to drop impact test conditions. So we have

\begin{align}\label{eq:varphi_G}
\varphi_{G^{\prime}} = \frac{G^{\prime}\left(\omega; \tau_{\text{age}} \approx 600~\text{s} \right)}{G^{\prime}\left(\omega; \tau_{\text{age}} \approx 10~\text{s} \right)}.
\end{align}

Now that $\varphi_{G^{\prime}}$ is determined, one needs $\varphi_{\gamma_{\text{y}}}$.  The yield strain is obtained from LAOS tests, data shown in Fig.~\ref{fig:GtoYSaging_YS}.  Viscoelastic moduli ($G^{\prime}$, $G^{\prime\prime}$) data for the three concentrations have a cross-over point. Conventionally, this is called the ``absolute yield strain''. One can also use the value of $\gamma_0$ for which $G^{\prime\prime}$ has a maxima, and this is called the ``dynamic yield strain''. Coincidentally, these two quantities for the samples used, both aged and unaged, were almost the same, $\gamma_{\rm y}\approx25\%$. This is consistent with results in the literature; yielding seems to occur at an approximately constant yield strain for colloidal dispersions \cite{MewisSpaull1976,Bonn_YS_JNNFM2016}. This gives

\begin{align}\label{eq:varphi_gamma}
\varphi_{\gamma_{\text{y}}} = \frac{\hat{\gamma}_{\text{y}}}{\gamma_{\text{y}}} \approx 1.
\end{align}
Using these results, we can now write

\begin{align}\label{eq:hypothesis_onlyYS_varphi_YS_repeat}
\varphi_{\sigma_{\rm y}} \equiv \varphi_{G^{\prime}} \varphi_{\gamma_{\text{y}}} \approx \varphi_{G^{\prime}},
\end{align}
where $\varphi_{G^{\prime}}$ is determined using Eq.~\ref{eq:varphi_G} and the aging data shown in Fig.~\ref{fig:laponiteflow}(B) (or equivalently, Fig.~\ref{fig:GtoYSaging_YS}(B), since $\gamma_{\text{y}}\approx25\%$ for all samples).

Using these results, the expression for the dimensionless group in Eq.~\ref{eq:eqn_IF_4} can be modified as

\begin{align}\label{eq:IF_hypothesis_onlyYS}
\text{IF}\left( \frac{D}{t} \right) \equiv \frac{\rho V^2 D}{\left[ \varphi_{G^{\prime}} \sigma_{\rm y} + K(V/t)^{0.5} + \eta_{\infty} V/t \right]t}.
\end{align}

\begin{figure}[tb]
  \begin{minipage}[!ht]{0.32\textwidth}
    \centering
        \includegraphics[scale=0.30,trim={0 0 0 0},clip]{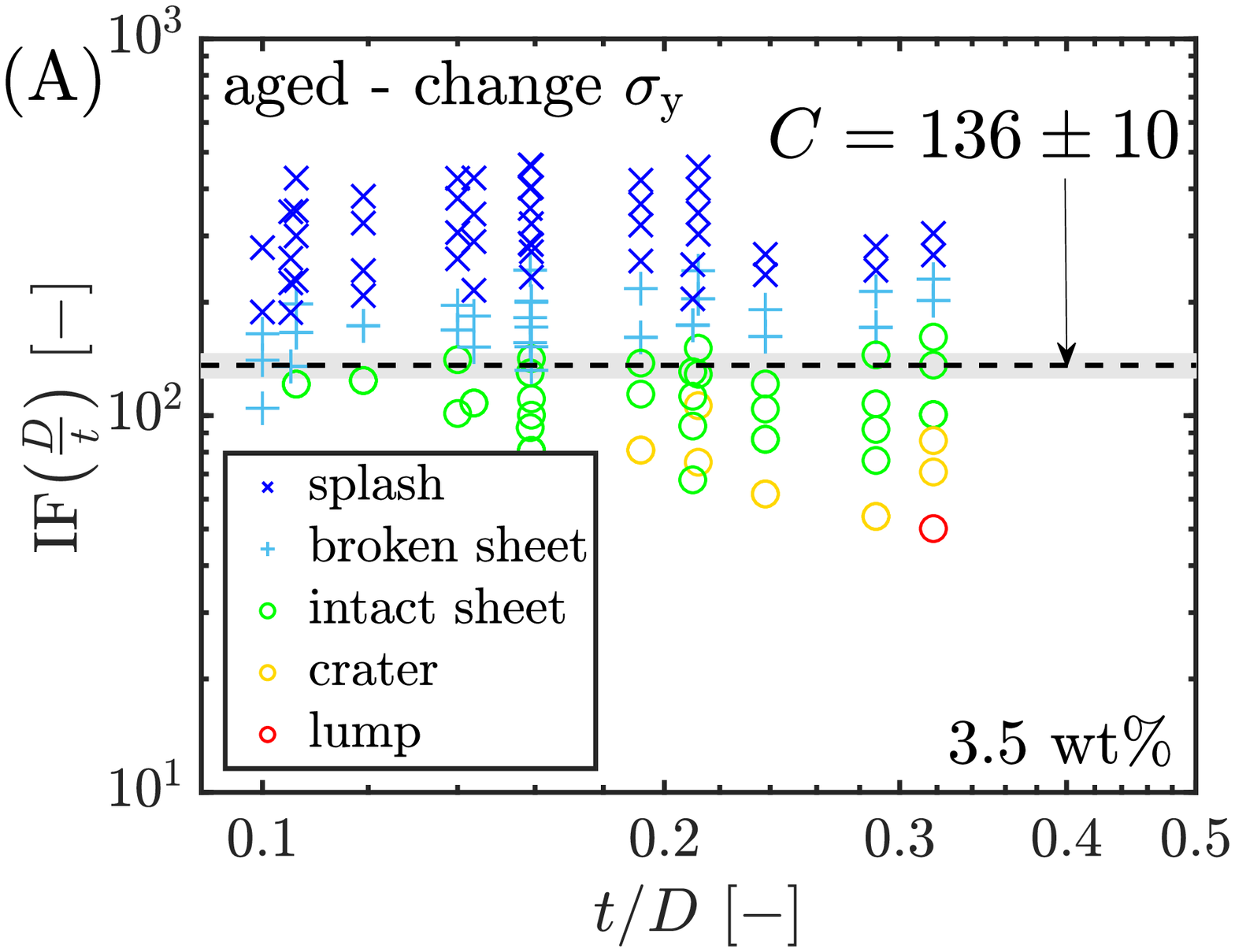}
  \end{minipage}
  \begin{minipage}[!ht]{0.32\textwidth}
    \centering
     \includegraphics[scale=0.30,trim={0 0 0 0},clip]{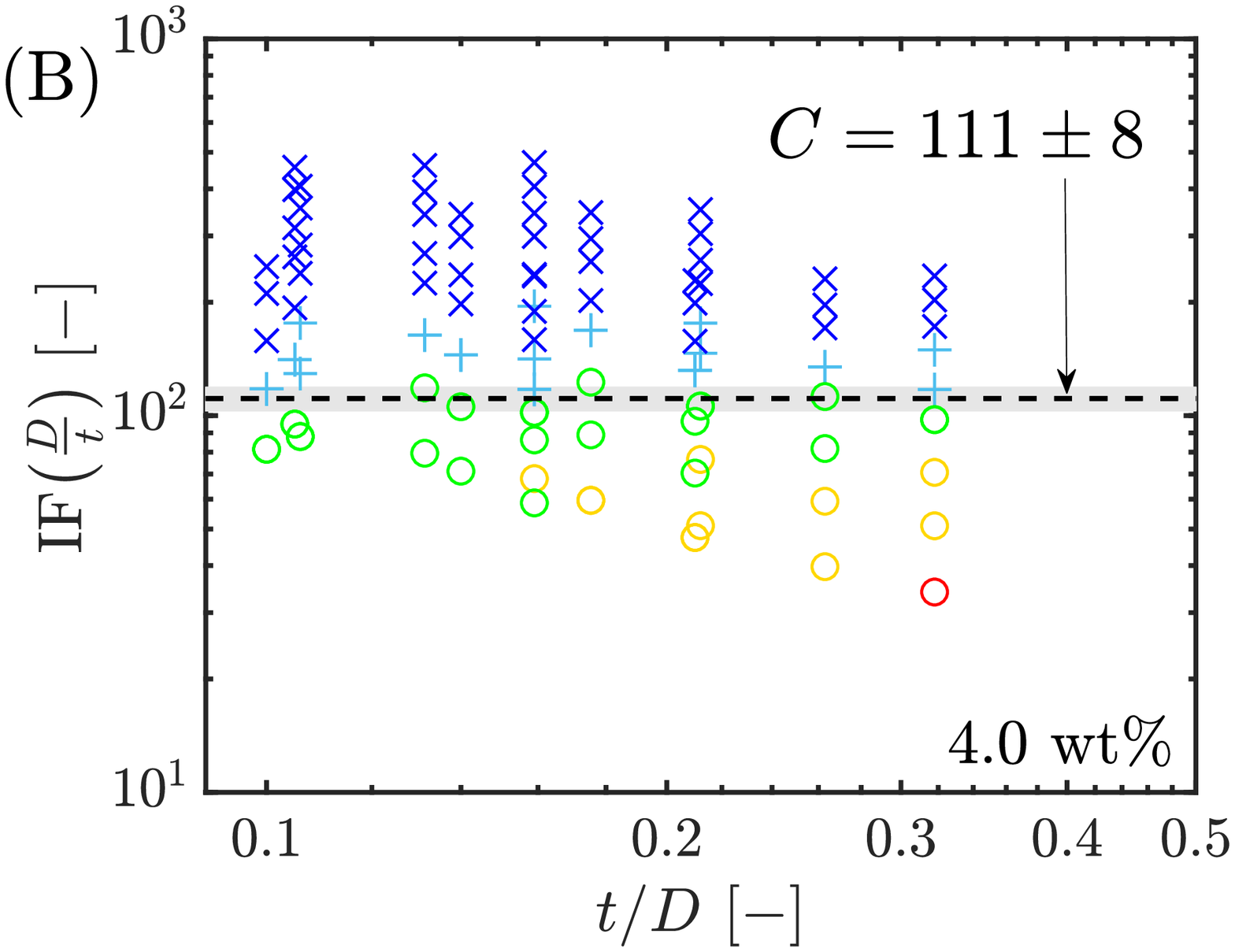}
  \end{minipage}
  \begin{minipage}[!ht]{0.32\textwidth}
    \centering
     \includegraphics[scale=0.30,trim={0 0 0 0},clip]{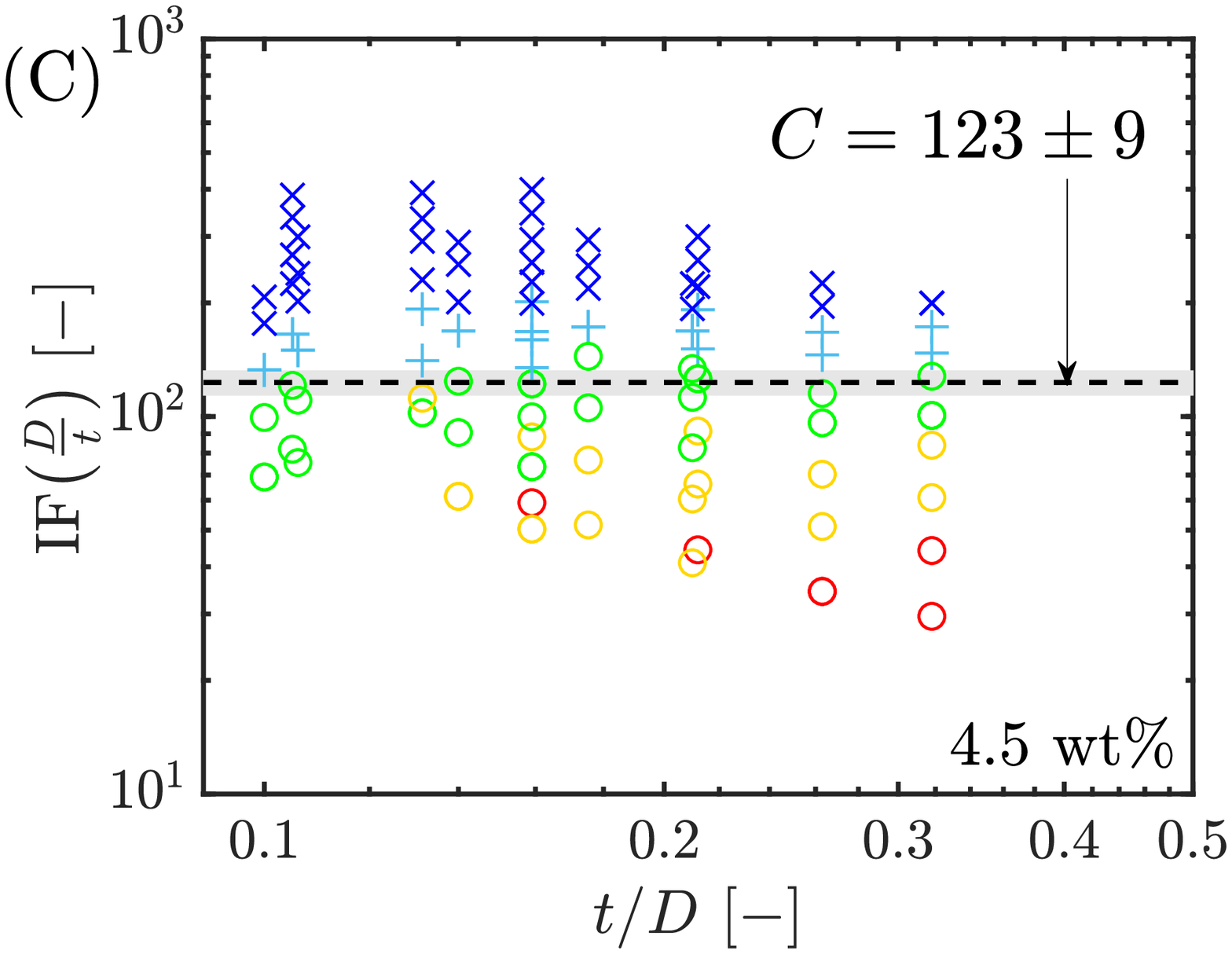}
  \end{minipage}
 \caption{Hypothesis 1 from Table~\ref{table:hypotheses_summary}: regime maps for aged Laponite using $G^{\prime}(\tau_{\text{age}})$ to update only $\sigma_{\rm y}$, Eq.~\ref{eq:IF_hypothesis_onlyYS}. The dimensionless group IF$(D/t)$ retains the same form, but with increased $\sigma_{\rm y}$ (and unchanged $K$, $\eta_{\infty}$).\label{fig:regimemaps_aged_GtoYS_YS}}
\end{figure}

We use this modified form of the dimensionless group to plot the regime maps for aged samples, shown in Fig.~\ref{fig:regimemaps_aged_GtoYS_YS}. From the new maps, we see that in addition to (i) good separation between the regimes and (ii) giving a constant $C$ for stick-splash transition, now the values of $C$ are (iii) comparable between the different concentrations of aged Laponite, and (iv) these values are similar to those for the unaged samples within experimental uncertainties. The values of $C$ for the  samples were $145,119,130$ for the unaged, compared to $136,111,123$ (previously 182, 150, and 155 before the shift using Eq.~\ref{eq:hypothesis_onlyYS_1},~\ref{eq:hypothesis_onlyYS_2}) for the aged, for 3.5, 4.0 and 4.5~wt\% respectively. It is surprising that such a simple modification should work.

\subsection{Other possible hypotheses\label{subsec:otherhypotheses}}

The hypothesis presented in the previous section was found to be the most credible among several possible approaches to account for thixotropic aging. All hypotheses considered are summarized in Table~\ref{table:hypotheses_summary}, and detailed in the SI. We describe them briefly here.

\begin{table}
\caption{Comparison of the performance of each hypothesis for predicting properties of aged samples, validated against the 4 criteria laid out in the text. The ``zeroth'' hypothesis is the control case: using the same steady shear properties for both unaged and aged samples, shown in Fig. \ref{fig:regimemaps_aged_uncorr}. For each hypothesis, the ``$+$'' and ``$-$'' symbols denote if the criteria are satisfied or not, respectively.\label{table:hypotheses_summary}}
  \begin{center}
\def~{\hphantom{0}}
 \begin{tabular}{P{1cm}|P{5cm}|P{1cm}P{1cm}P{1cm}P{1cm}}
 \multicolumn{2}{P{6cm}|}{\multirow{2}{*}{Details of each hypothesis}} & \multicolumn{4}{P{4cm}}{Validity criteria}\\
 \cline{3-6}
 \multicolumn{2}{P{1cm}|}{} & \multicolumn{1}{P{1cm}}{i} & \multicolumn{1}{|P{1cm}}{ii}  & \multicolumn{1}{|P{1cm}}{iii} & \multicolumn{1}{|P{1cm}}{iv} \\
 \hline
 
 \multirow{3}{*}{0.} & $\varphi_{\sigma_{\rm y}} = 1$ & \multirow{3}{*}{$+$}  & \multirow{3}{*}{$+$} & \multirow{3}{*}{$-$} & \multirow{3}{*}{$-$} \\
 & $\varphi_{K} = 1$ & \\
 & $\varphi_{\eta_{\infty}} = 1$ & \\
\hline

\multirow{3}{*}{1.} & $\;\;\;\;\;\;\;\;\;\varphi_{\sigma_{\rm y}} = \varphi_{G^{\prime}} \varphi_{\gamma_{\text{y}}}$ & \multirow{3}{*}{$+$}  & \multirow{3}{*}{$+$} & \multirow{3}{*}{$+$} & \multirow{3}{*}{$+$} \\
& $\varphi_{K} = 1$ & \\
 & $\varphi_{\eta_{\infty}} = 1$ & \\
\hline

 \multirow{3}{*}{2.} & $\;\;\;\;\;\;\;\;\;\varphi_{\sigma_{\rm y}} = \varphi_{G^{\prime}} \varphi_{\gamma_{\text{y}}}$ & \multirow{3}{*}{$+$}  & \multirow{3}{*}{$+$} & \multirow{3}{*}{$-$} & \multirow{3}{*}{$-$} \\
 & $\;\;\;\;\;\;\;\;\varphi_{K} = \varphi_{G^{\prime}} \varphi_{\gamma_{\text{y}}}$ & \\
 & $\;\;\;\;\;\;\;\;\varphi_{\eta_{\infty}} = \varphi_{G^{\prime}} \varphi_{\gamma_{\text{y}}}$ & \\
\hline

 \multirow{3}{*}{3.} & $\;\;\;\;\;\;\;\,\sigma_{\text{peak}} = \Sigma + \kappa\dot{\gamma}^{0.5} + H\dot{\gamma}$ & \multirow{3}{*}{$-$}  & \multirow{3}{*}{$+$} & \multirow{3}{*}{$+$} & \multirow{3}{*}{$+$} \\
 & $\;\;\:\hat{\sigma}_{\text{y}} = \Sigma$ & \\
 & $\;\;\;\;\;\;\;\;\;\;\hat{K} = \kappa$, $\hat{\eta}_{\infty} = H$ & \\
\end{tabular}
  \end{center}
\end{table}

Hypothesis ``0'' is the control case, where the same steady state properties are used for both unaged and aged samples. Thus, $\varphi_{\sigma_{\rm y}}=1$, $\varphi_{K}=1$, $\varphi_{\eta_{\infty}}=1$ in Eq.~\ref{eq:eqn_IF_4}. As shown in Fig.~\ref{fig:regimemaps_aged_uncorr}, it works well to separate the impact types and also gives a constant regime boundary for stick-splash transition (criteria i and ii). But the values of $C$ are not similar for different Laponite concentrations, with large differences between $C$ for aged and unaged fluids (fails to satisfy criteria iii and iv). This result might be expected, as evident from the video stills in Fig.~\ref{fig:velocityvaried}. But its failure does give insight about breakdown timescales, indicating that these must be sufficiently longer than drop impact timescales as discussed in Sec.~\ref{subsec:dropimpacts}.

Next, Hypothesis 1 is where $\varphi_{\sigma_{\rm y}} = \varphi_{G^{\prime}}$, and $\varphi_{K} = \varphi_{\eta_{\infty}} = 1$ in Eq.~\ref{eq:eqn_IF_4}, already explained in detail in Sec.~\ref{subsec:agedprops_hypothesis_onlyYS}, and results shown in Fig.~\ref{fig:regimemaps_aged_GtoYS_YS}. It is found to be the most credible, satisfying each of the four criteria.

Hypothesis 2 is a more natural and obvious hypothesis, to increase each of $\sigma_{\rm y}$, $K$ and $\eta_{\infty}$ by the same factor $\varphi_{G^{\prime}}$ (discussed in detail in the SI, Sec.~2). This follows the most common structure-parameters models in thixotropy which predict that all flow properties should increase with structure growth \cite{MewisWagnerReview2009,LarsonReview2019}. That is, $\varphi_{\sigma_{\rm y}} = \varphi_{K} = \varphi_{\eta_{\infty}} \equiv \varphi_{G^{\prime}}$ in Eq.~\ref{eq:eqn_IF_4}. But our results suggest otherwise. The overlap (blurring of boundaries) between different impact regimes increased, which is quantified by the increase in uncertainty values for $C$ (Fig.~S4). Although the regime boundary is of an approximately zero slope, giving a constant value of $C$, the values are different for different Laponite concentrations, and outside the uncertainty. They also do not compare well with the unaged $C$ values, in that the mean values are significantly different for there to be no overlap even within the uncertainties.

Finally, Hypothesis 3 changes each of $\sigma_{\rm y}$, $K$, and $\eta_{\infty}$ with aging, but now the aged properties use peak stresses during startup of shear tests (e.g.\ Fig.~\ref{fig:laponiteflow}(A), inset), since this transient test may map more closely the short timescales of to the drop impact event \cite{KochPhDThesis}. The peak stresses are used as a measure of a rate-dependent static yield stress and a rate-dependent $K$ and $\eta_{\infty}$. The peak stress for each test is plotted versus the corresponding rate, and the plot is assumed to represent a flow curve, to which Eq.~\ref{eq:GenHB} is fitted. The fit parameters now give the aged $\sigma_{\rm y}$, $K$, and $\eta_{\infty}$ (renamed $\Sigma$, $\kappa$, and $H$ respectively, shown in Table~\ref{table:hypotheses_summary}). As a sample ages and structures build up, the overshoot stresses should also increase because the sample gets stiffer and more viscous. So this technique reflects the state of aging of a sample, including rate dependence. This method works well in all aspects except the most important one: the separation between regimes is lost, with very significant overlap, as seen from the uncertainties on $C$ which are comparable to the mean $C$ values. So this is not the most suitable method (Figs.~S7, S8, S9). Detailed description of this method is in the SI, Sec.~3.

\section{Conclusions\label{sec:conclusions}}
We report the first-ever study of thixotropic effects in viscoplastic drop impact, generating a large systematic data set covering 1008 different impact conditions across a range of concentration with aqueous Laponite suspensions. We focused on classification of the type of impact and whether material is retained (sticks) or is ejected away (splashes) at the impact site. 
Aged materials have less splash, all else being equal. This reveals that thixotropic breakdown timescales are long enough to affect the dynamics. 
All impact conditions are collapsed into a dimensionless group, IF$(D/t)$, originally proposed in~\cite{BCB_PhysFluids2015, SSRHE_JFM2020}. 
The critical value demarcating the boundary between ``stick'' and ``splash'' types is larger for aged compared to unaged samples when the same steady shear properties are used for both states of aging.

We propose and test hypotheses for including thixotropic aging in the Herschel-Bulkley parameters of the dimensionless group $\text{IF}(D/t)$. The hypothesis with the strongest evidence is to increase the dynamic yield stress $\sigma_{\rm y}$ proportional to the aged linear elastic modulus $G^{\prime}$, while the parameters for rate-dependent effects, $K$ and $\eta_{\infty}$, remain unchanged in spite of aging. This is surprising, as one might instead expect all flow parameters to increase with structure build-up as proposed in some theoretical thixotropy models \cite{MewisWagnerReview2009,LarsonReview2019}. But droplet impact timescales are short ($\mathcal{O}(10)~\rm{ms}$), and $\sigma_{\rm y}$ may break down sufficiently slowly (following the argument presented in Sec.~\ref{subsec:dropimpacts}, Fig.~\ref{fig:laponiteflow}(A), inset) to dominate dissipative effects in aged samples compared to $K$ and $\eta_{\infty}$. 
Our observations imply the dominance of the plastic component of the total dissipative stress during impact.
This hypothesis of increasing only $\sigma_{\rm y}$ satisfies all four criteria for testing efficacy of a dimensionless group,
unlike other hypotheses which were found to be less supported, as summarized in Table~\ref{table:hypotheses_summary}.

Effects including surface tension and extensional rheology properties are neglected for these test conditions. These all seem to be valid assumptions for the fluids and conditions considered in our work here. The scaling law might fail if we test fluids with significant extensibility (which can be significant and even desirable for certain yield-stress fluids \cite{Nelson2018,Rauzan2018}), or probe scenarios where surface tension effects become comparable with viscous and inertial stresses, e.g.\ at much smaller drop sizes which are difficult to produce in our setup (and may need to be studied as spray rather than isolated individual drops). Extreme values of the normalized coating thickness ($t/D$) may also make the situation more complicated, necessitating the use of additional dimensionless groups instead of the single one of Eq.~\ref{eq:IFintro}.

Our study focused on different thixotropic \textit{aging} of samples, but thixotropic \textit{breakdown} timescales were also considered to interpret the results. Dimensionless timescales clarify the range of conditions. During aging, the dimensionless thixotropic recovery time $\mathcal{T}_+$ ranges from very small ($\mathcal{T}_+\ll 1$, unaged) to order one ($\mathcal{T}_+ \sim\mathcal{O}(1)$, aged), defined as  $\mathcal{T}_+\equiv\frac{\tau_{\rm{exp}}}{\tau_{\rm{thixo},+}}$, with experimental time $\tau_{\rm{exp}}=\tau_{\rm{age}}$ and characteristic thixotropic recovery time $\tau_{\rm{thixo},+}$.
This comes from Fig.~\ref{fig:laponiteflow}(B) which shows that samples continue to recover/age significantly at 600~s and the process is not yet approaching a steady state. 
In contrast, during the short time droplet impact ($\tau_{\rm{exp}}\sim\mathcal{O}(10)$~ms), recovery has no chance to occur ($\mathcal{T}_+\ll 1$). Breakdown is more nuanced, with characteristic breakdown time scale $\tau_{\rm{thixo},-}\sim\mathcal{O}(100)$~ms (Fig.~\ref{fig:laponiteflow}(A), inset), and therefore $\mathcal{T}_-\sim\mathcal{O}(0.1)$, where 
$\mathcal{T}_-\equiv\frac{\tau_{\rm{exp}}}{\tau_{\rm{thixo},-}}$. Having $\mathcal{T}_-<1$ indicates that structure is not fully broken down by the impact. 
This is why aged properties can play a significant role, and why these viscoplastic fluids can still ``stick'' at the impact location with a finite yield stress even after being subjected to large deformation rates.
Fluids with more fragile particulate networks may be different, if they have faster breakdown timescale $\tau_{\rm{thixo},-}$ compared to droplet impact. In that case, aging effects are quickly removed and negligible, and structure breakdown might make it difficult for the fluid coating to be retained. 
These are important considerations in applications such as spray coating on vertical surfaces, which involve problems of draining of coating, where the subsequent thixotropic recovery ($\tau_{\rm{thixo},+}$) is important for long timescale coating retention.

The group tested here, in Eq.~\ref{eq:IFintro} for unaged and Eq.~\ref{eq:eqn_IF_4} for aged, provides fundamental insight and will be very useful in predicting splash behavior for thixotropic yield-stress fluids in a variety of applications from spray coating to fire suppression \cite{truscott,truscott_personal,appel2016,BCBPhDThesis}. We have also observed this qualitative aging effect in more complex flow scenarios for drop impact into porous structures (meshes) \cite{SSRHE_GFM2018}; the effect is perhaps even more dramatic and relevant to fire suppression and coating complex substrates.

\begin{acknowledgements}
This work was funded by the National Science Foundation, CAREER Award, CBET-1351342. S.S. acknowledges the PPG Foundation for partial support through a PPG-MRL Graduate Research Assistantship at the University of Illinois at Urbana-Champaign. S.S. also thanks D. Bonn, Y. Wang and N. Ramlawi for their insightful suggestions.
\end{acknowledgements}

\bibliography{02-Manuscript-drop-impact-Laponite-aged}

\end{document}